\documentclass{aa}
\usepackage{graphicx}      
\usepackage{txfonts}
\usepackage{natbib}
\usepackage{epsfig}
\usepackage{subfigure}
\bibpunct{(}{)}{;}{a}{}{,} 

\begin{document}
   \title{The excitation of near-infrared H$_2$ emission in NGC 253}

   \author{M.~J.~F.~Rosenberg
          \inst{1}
          \and
          P.~P. van der Werf \inst{1}
          \and
        F.~P. Israel \inst{1}
          }

   \institute{Sterrewacht Leiden, Universiteit Leiden,
              P.O. Box 9513, NL-2300 RA Leiden\\
              \email{rosenberg@strw.leidenuniv.nl}
             }

   \date{Accepted Dec 6th 2012}

   \abstract
   {Because of its large angular size and proximity to the Milky Way, NGC 253, an archetypal starburst galaxy, provides an excellent laboratory to study the intricacies of this intense episode of star formation.}
   {We aim to characterize the excitation mechanisms driving the emission in NGC 253.  Specifically we aim to distinguish between shock excitation and UV excitation as the dominant driving mechanism, using Br$\gamma$, H$_2$ and [FeII] as diagnostic emission line tracers.}
   {Using SINFONI observations, we create linemaps of Br$\gamma$, [FeII]$_{1.64}$, and all detected H$_2$ transitions. By using symmetry arguments of the gas and stellar gas velocity field, we find a kinematic center in agreement with previous determinations.  The ratio of the 2-1 S(1) to 1-0 S(1) H$_2$ transitions can be used as a diagnostic to discriminate between shock and fluorescent excitation. }
   {Using the 1-0 S(1)/2-1 S(1) line ratio as well as several other H$_2$ line ratios and the morphological comparison between H$_2$ and Br$\gamma$ and [FeII], we find that excitation from UV photons is the dominant excitation mechanisms throughout NGC 253.  We employ a diagnostic energy level diagram to quantitatively differentiate between mechanisms.  We compare the observed energy level diagrams to PDR and shock models and find that in most regions and over the galaxy as a whole, fluorescent excitation is the dominant mechanism exciting the H$_2$ gas. We also place an upper limit of the percentage of shock excited H$_2$ at 29\%.} 
   {We find that UV radiation is the dominant excitation mechanism for the H$_2$ emission.  The H$_2$ emission does not correlate well with Br$\gamma$ but closely traces the PAH emission, showing that not only is H$_2$ fluorescently excited, but it is predominately excited by slightly lower mass stars than O stars which excite Br$\gamma$, such as B stars.}

\titlerunning{H$_2$ excitation in NGC 253}
\authorrunning{Rosenberg, M.~J.~F. et al.}

   \maketitle
%

\section{Introduction}

Understanding the excitation mechanisms and drivers in nearby starburst galaxies is critical to understanding their role in galaxy evolution. One of the best laboratories to study this is NGC 253.  This galaxy is a nearby (3.5 Mpc \cite{2005MNRAS.361..330R}), edge-on infrared bright galaxy that is part of the Sculptor group.  It is classified as a late-type barred spiral galaxy (SAB(s)c) with star formation confined to the inner 0.5 kpc. The star formation rate of the starburst is 1.4-9.5 M$_{\odot}$ yr$^{-1}$ \citep{2003MNRAS.339..793G}.  This intense episode of star formation is driven by a 7 kpc bar, which funnels gas into the nucleus \citep{1998ApJ...505..639E}.  There is a starburst driven superwind extending perpendicular to the disk \citep{1970ApJ...159..799D,1978ApJ...219..424U,1984ApJ...286..491F,1987AJ.....93..264M,2011MNRAS.414.3719W}.  

The inner region of star formation is highly obscured with an A$_V$ of $\sim 5-18$ mag \citep{2009ApJ...697.1180K,2009MNRAS.392L..16F}.  Many dense, luminous, dust enshrouded young stellar clusters have been found in the nucleus \citep{1993ApJ...406L..11F,2000MNRAS.312..689F}.  In addition to the stellar clusters, \citet{2003AJ....125.1210A} and \citet{1997ApJ...488..621U} have observed many compact radio sources in [FeII] and radio respectively, representing young supernova remnants. \citet{1985ApJ...299L..77T} observed over 60 compact radio sources, the brightest of which is named TH2.  A high resolution (1''-2'') study of the molecular gas in NGC 253 was performed by \citet{2011ApJ...735...19S}.  They found 5 warm, dense clumps of molecular gas, which are all coincident with radio sources.  They suggest that these clumps are natal molecular cloud complexes harboring massive star formation.  The K band continuum peak, which was originally defined as the nucleus by \citet{1991ApJ...380L..63F}, is the 
location of a super star cluster.  This 
super star cluster was observed with Hubble Space Telescope by \citet{1996AJ....112..534W} and is consistent with a single stellar population.  In the mid infrared, two bright peaks have been observed in the central regions.  The brightest of the two is coincident with the K band continuum peak and super star cluster and has no radio counterpart \citep{1997ApJ...488..621U}.  The second, 3.0'' to the northeast is coincident to the radio peak TH2.  A more recent study by \citet{1997ApJ...488..621U} suggests that TH2 is the true nucleus of NGC 253 and represents either a low-luminosity AGN or a compact supernova remnant.   

Although TH2 is generally considered to be the nucleus, a recent study by \citet{2010ApJ...716.1166M} provides new estimates for the kinematic center based on the kinematics of the stellar component.  Using SINFONI on the VLT, they were able to achieve very high angular resolution and derive the stellar velocity field.  They found the kinematic center to be offset from TH2 only by $\Delta$x=0''.6 and $\Delta$y=0''.4 (i.e. by 12 pc) in the southwest direction.  They propose an alternative position for the kinematic center as the X-ray source X-1 \citep{2002ApJ...576L..19W}.  Although NGC 253 has been the target for many detailed kinematic studies, there is still much uncertainty about the precise location of the true kinematic center.

The subject of excitation mechanisms has been extensively studied in NGC 253.  \citet{2006ApJS..164..450M} finds that the chemistry and heating of NGC 253 is dominated by large scale, low velocity shocks.  Presence of shocked molecular material is indeed evident through the presence of widespread SiO emission throughout the nuclear regions \citep{2000A&A...355..499G}. \citet{2009ApJ...706.1323M} suggest that, although NGC 253 is dominated by shock chemistry, PDRs play a crucial role in the chemistry, since there are very high abundances of PDR tracing molecules, namely HCO$^+$, CO$^+$.  In addition, they find that although NGC 253 was thought to be at an earlier stage of evolution than M 82, a prototypical starburst with strong PDR characteristics, the molecular clouds are larger and have a higher column density, pointing to a later stage of evolution, such as seen in M 82.  \citet{2009ApJ...694..610M} also find that the UV heating in NGC 253 is more similar to M 82 than other starbursts, yet still 
not the dominant excitation mechanism.  \citet{1998MNRAS.297..624H} used the ortho- to para- ratio of H$_2$, to argue that most of the H$_2$ is excited by PDRs instead of shocks.  In addition, \citet{1989MNRAS.236...89I,1994ApJ...436L.185L} show in NGC 604 and Orion A respectively that although in small-beam apertures it may seem as if high surface brightness H$_2$ emission from shocks is dominant, when the beam size increases, it is clear that fluorescent excitation dominates on large scales.   

In this paper, we investigate the excitation mechanisms for the molecular gas in the nuclear region of NGC 253.  We use the following methods to differentiate between shock dominated and PDR dominated excitation on a 2 pc scale:
\begin{itemize}
 \item Calculate the 2-1 S(1)/1-0 (S1) line ratio at each pixel position
 \item Compare morphology of Br$\gamma$, [FeII], and PAHs to that of the H$_2$
 \item Relate observed H$_2$ excitation diagrams to those generated by shock and PDR models.
\end{itemize}

In Section~\ref{sec:obs}, we will discuss the observations, while in Section~\ref{sec:res} we present the derived spectra and linemaps.  In Section~\ref{sec:kin} we will investigate and define the kinematic center.  Then, in Section~\ref{sec:h2}, we investigate the nature of the dominant excitation mechanisms throughout the galaxy, focusing on the dense clumps.  In Section~\ref{sec:mod}, we compare our excitation diagrams to shock and PDR models spanning the parameter space.  We summarize our conclusions and their implications in Section~\ref{sec:conc}.

\section{Observations}
\label{sec:obs}
\subsection{SINFONI Observations}
All observations were made with the Spectrograph
for INtegral Field Observations in the Near-Infrared (SINFONI) at the
ESO VLT.  SINFONI provides spatial and spectral data in the form of data
cubes in J, H, and K bands.  The SINFONI instrument is mounted at the
Cassegrain focus of the Unit Telescope 4 at the Very Large Telescope
(VLT).

We observed in the H, and K bands using a spatial
pixel scale of 0.25'' corresponding to a field of view of
8''$\times$8'' per frame and a spectral resolution of 2000, 3000 and 4000
respectively, which corresponds to a velocity resolution of 149.8, 99.9 and 74.9 km/s.  All science
observations were taken in the ABA'nodding mode (300s of object, 300s
of sky, 300s of object), where A' is slightly offset from A.  The
object exposures are aligned and averaged during the reconstruction of the data
cube.

The observations of NGC 253 were made in visitor mode on August 28th, 2005.  In order to capture the full extent of the H$_2$ emission, consecutive frames were taken in the K band moving further away from the center, along the disk until H$_2$ was no longer detected.  This resulted in 6 separate pointings.  Since there are also H$_2$ transitions in the H band, a similar strategy was used, resulting in 4 separate pointings.  

We used the standard reduction techniques of the SINFONI pipeline on
all observations. including corrections for flat field, dark current,
nonlinearity of pixels, distortion, and wavelength calibration.  We
obtained the flux calibration and atmospheric corrections from
observations of a standard star, namely HR 2058 in the H band and HD 20001 in the K band.  This is the same dataset \citet{2010ApJ...716.1166M} used to determine the kinematic center.
\subsection{ISAAC Observations}
We use the ISAAC observations of NGC253 in the 3.21 and 3.28 $\mu$m filters described by Tacconi-Garman et al. (2005). These observations were uncalibrated. We established a flux calibration for these two images using the ISO-SWS spectra by Sturm et al (2000), taking into account the filter profiles of the two ISAAC filters and the precise location and orientations of the ISO-SWS aperture. Inspection of the spectrum by Sturm et al (2000) shows that the 3.21 $\mu$m filter contains only continuum emission, while the 3.28 $\mu$m filter contains continuum+PAH emission.  The peak of the PAH emission is at 3.3 $\mu$m, which is fully included in the 3.28 $\mu$m filter. The spectral slope is small enough that there is negligible change in continuum level between 3.21 and 3.28 $\mu$m and therefore a PAH emission image was created by subtracting the calibrated 3.21 $\mu$m image from the calibrated 3.28 $\mu$m image.

\section{Results}
\label{sec:res}
\begin{figure*}
\centering
\includegraphics[width=15cm]{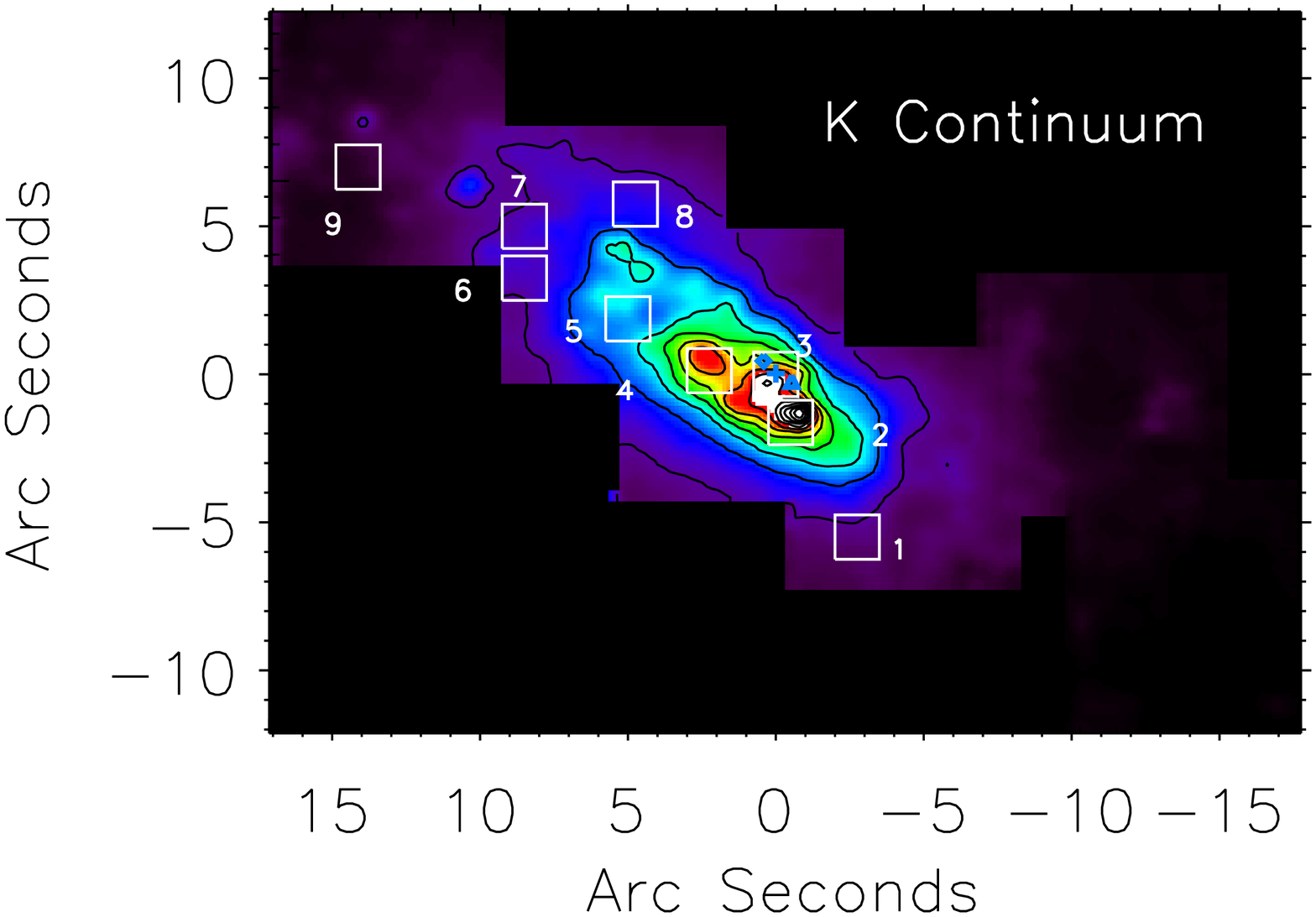}
\includegraphics[width=15cm]{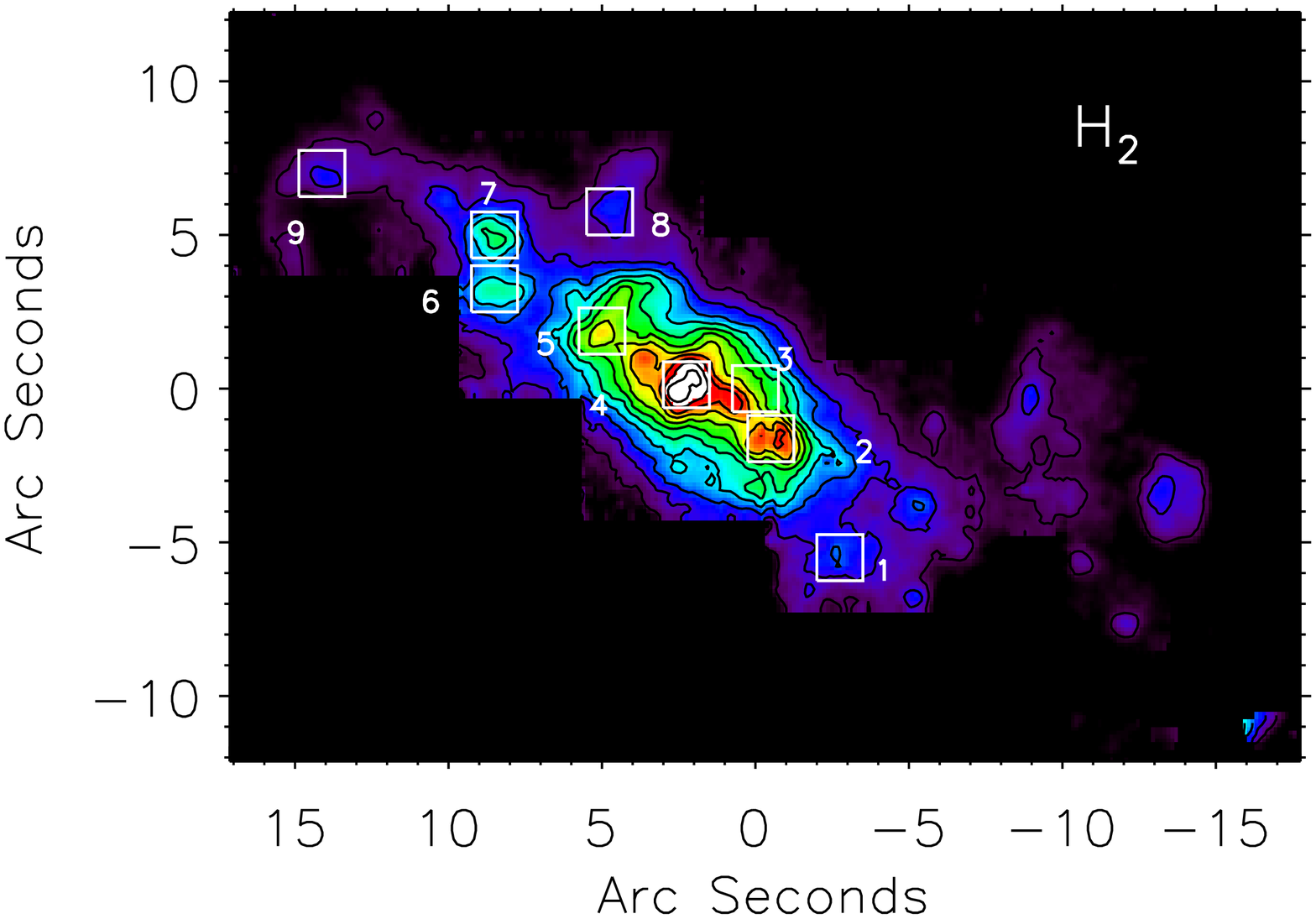}
\caption{K band continuum map and H$_2$ 1-0 S(1) line map of NGC 253. The offset from the kinematic center at position $(\alpha,\delta)_{2000}=(00^{h}47^{m}33^{s}.14,-25^{\circ}17'17''.52)$ is given on each axis.  The contours represent 13 equally spaced levels with a maximum flux of 1.3$\times 10^{-11}$ erg s$^{-1}$cm$^{-2}\mu m^{-1}$arcsec$^{-2}$ and $5.4\times10^{-15}$ erg s$^{-1}$cm$^{-2}$arcsec$^{-2}$,for the K band continuum and H$_2$ respectively.  The white rectangular regions are H$_2$ flux peaks, with the exception of Region 3, which is centered on the kinematic center.  The blue '+' represents the kinematic center, the blue triangle represents X-1 and the blue diamond represents TH2.}
\label{fig:kband}
\end{figure*}

\begin{figure*}
\centering
\includegraphics[width=15cm]{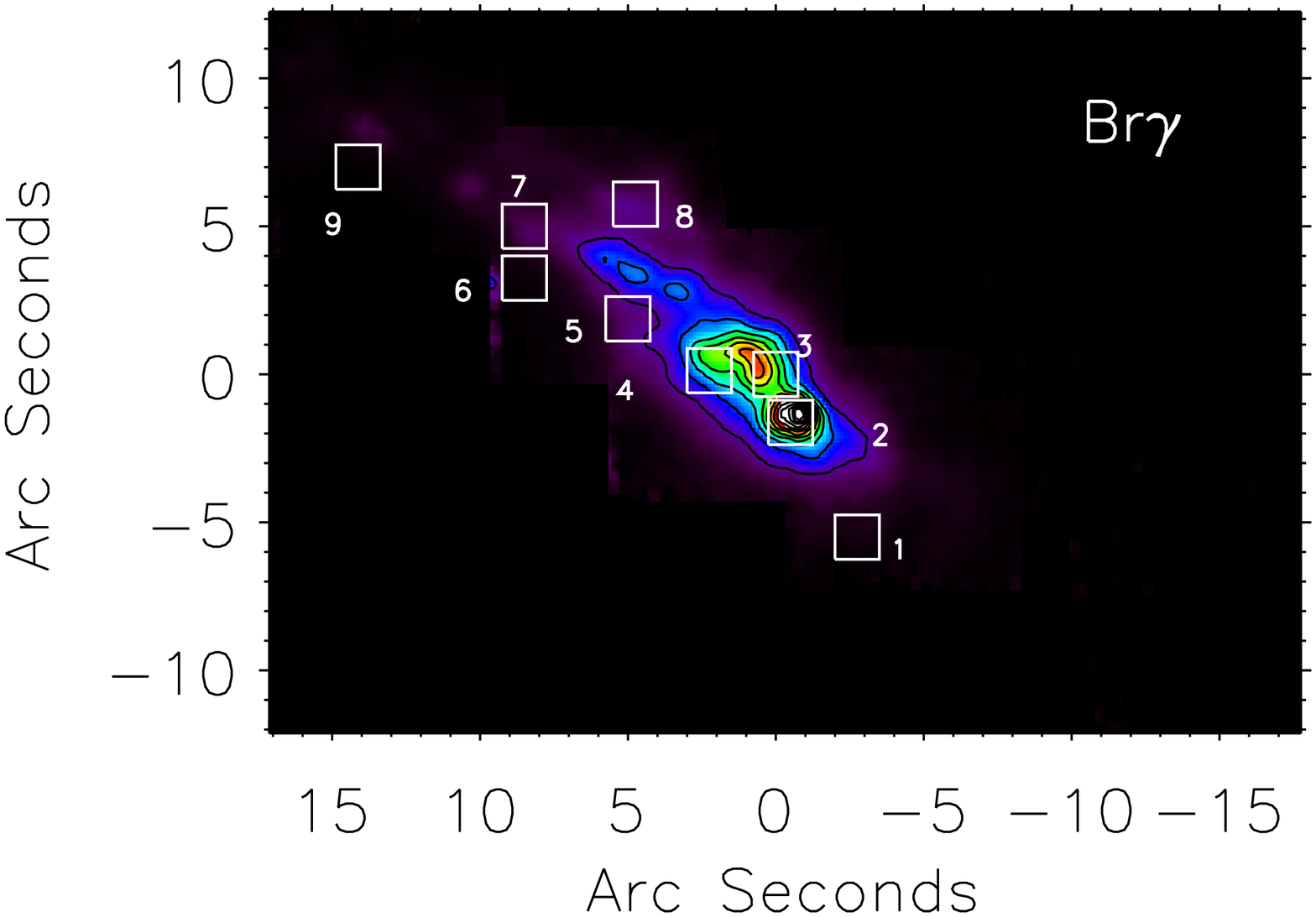}
\includegraphics[width=15cm]{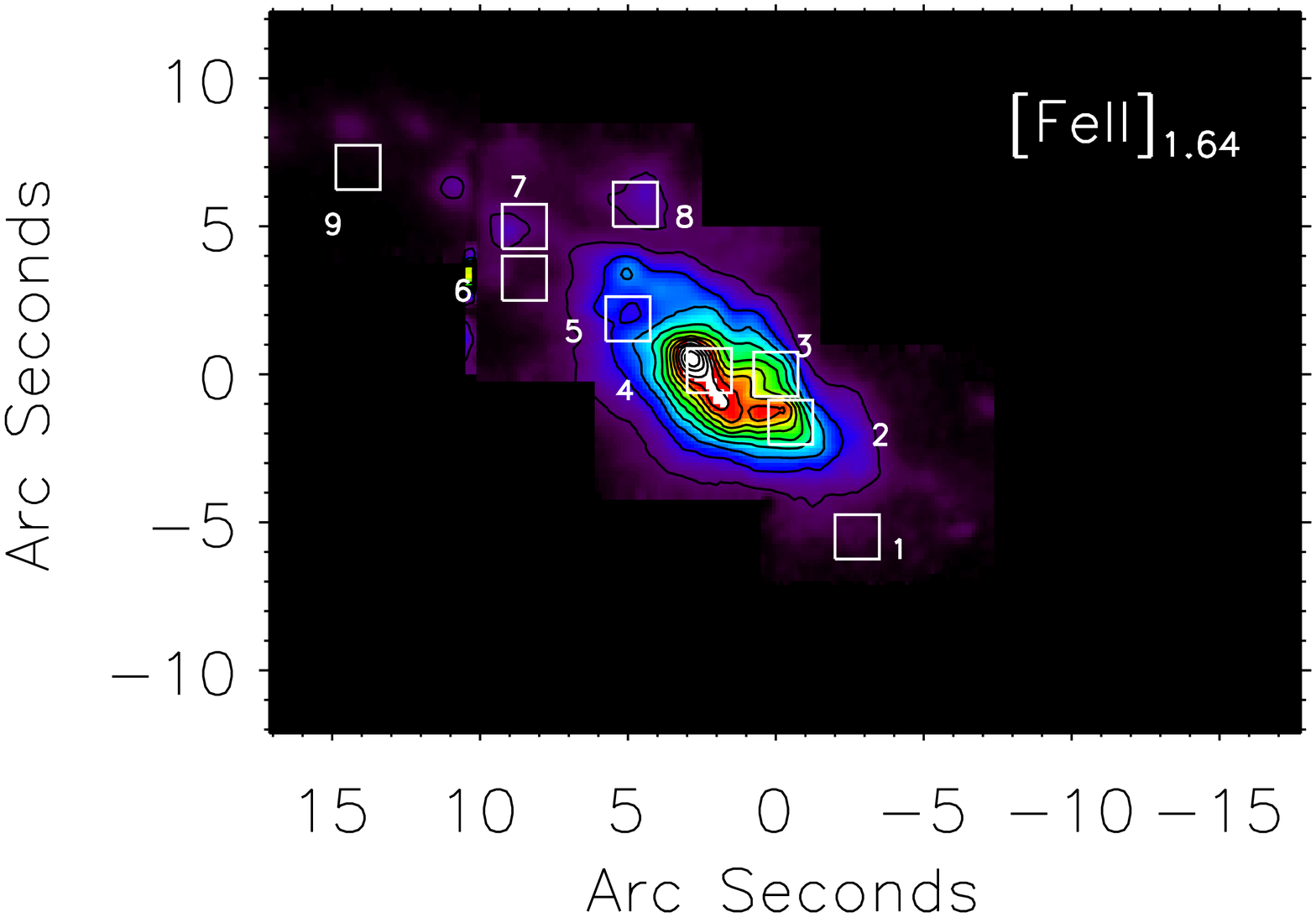}
\caption{Br$\gamma$ and [FeII]$_{1.64}$ emission line map of NGC 253. The offset from the kinematic center at position $(\alpha,\delta)_{2000}=(00^{h}47^{m}33^{s}.14,-25^{\circ}17'17''.52)$ is given on each axis. The contours represent 13 equally spaced levels with a maximum flux of 4.5$\times 10^{-14}$ and 2.8$\times 10^{-14}$ erg s$^{-1}$cm$^{-2}$arcsec$^{-2}$ for Br$\gamma$ and [FeII] respectively.  The white rectangular regions are areas of concentrated H$_2$ flux, with the exception of Region 3, which is centered on the kinematic center.}
\label{fig:bg}
\end{figure*}

The SINFONI datacubes allow us to construct continuum maps, line maps and velocity maps of emission lines detected in these
bands. The K band continuum map and H$_2$ 1-0 S(1) emission line map are shown in Figure~\ref{fig:kband} and the Br$\gamma$ and [FeII]$_{1.64}$ emission line maps are shown in Figure~\ref{fig:bg}.  All line maps are centered on the kinematic center proposed by \citet{2010ApJ...716.1166M}.  We have chosen 8 regions defined by the most dominant H$_2$ peaks and these regions are overplotted by white rectangles in each map. There is an additional region centered on the kinematic center.  All regions are 12$\times$12 pixels or 1.5$\times$1.5 arcseconds.  These regions were selected based on H$_2$ flux. There is a relatively bright H$_2$ emission clump to the north west of Region 1, which was not selected for analysis since it shows similar emission features to Region 1.     

%
%

The K band continuum map reveals the inclined disk of the galaxy, tracing the older stellar population.  The disk appears irregular with clear clumps of high luminosity.  The kinematic center (Region 3) shows a very high level of continuum emission. In addition, superimposed on the bright continuum emission around the kinematic center, there is a bright flux peak at the location of the super star cluster (Region 2) studied by \citet{1996AJ....112..534W}.  

On the other hand, the H$_2$ 1-0 S(1) map shows a completely different morphology. The H$_2$ extends along the entire 
disk with its brightest clump located several arcseconds east of the kinematic center (Region 3).  However, the H$_2$ is even more inhomogeneous than the K band continuum.  There are clear H$_2$ bright clumps ranging out into the furthest regions.  The white rectangles highlight the brightest emission clumps.  There is also significant H$_2$ emission originating from the super star cluster (Region 2), but not from the center.

Similarly, comparison of the H$_2$ and the Br$\gamma$ map reveals little correlation.  The Br$\gamma$ flux peak is coincident with the super star cluster, but shows no clear flux increase near Region 4, where H$_2$ is brightest, nor at the center.  The Br$\gamma$ is also generally less extended than the H$_2$, with most of the emission in the southwestern portion of the galaxy, centered on the super star cluster. 

The [FeII]$_{1.64}$ map also reveals emission peaking near, but not coincident with, Region 4. In the northeastern regions, the [FeII] emission is also quite clumpy, with some of the clumps coinciding with the H$_2$, such as Regions 5, 7 and 8.  Comparing the [FeII] to the Br$\gamma$ emission, we see in both cases, as well as in the K continuum, an elongated arm-like feature coming from the center, curving out towards the northeast corner along the disk, bordering on the northern edge of Region 5.  However the [FeII] feature is spatially offset towards the south with respect to the Br$\gamma$ feature, as noted originally by \citet{1993ApJ...406L..11F}.  The H$_2$ and continuum emission in this region traces neither Br$\gamma$ or [FeII].     

\begin{figure*}
\centering

 \includegraphics[width =17cm] {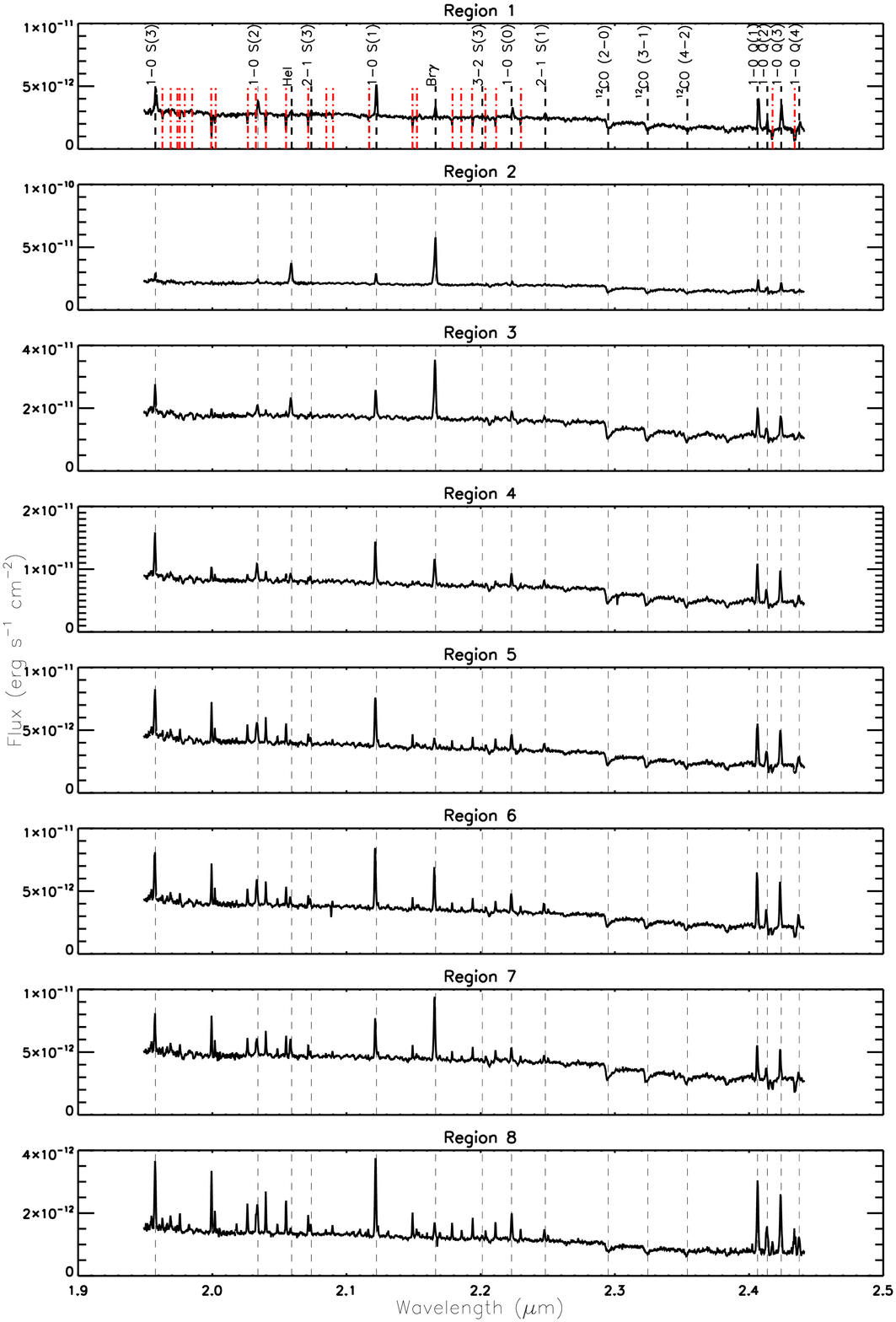}
   \label{fig:spec}
\caption{K band spectra of H$_2$ bright clumps.  The integrated K band spectra, in the rest frame, integrated over the regions shown in Figure~\ref{fig:kband}.  Vertical dotted lines are used to show dominant emission lines.  The strongest OH lines have been marked in the spectrum of Region 1 with the red dot-dash line.  The $^{12}$CO absorption bands are also visible at 2.294, 2.323, and 2.353 $\mu$m respectively.  Region 2 is the K Continuum peak and hot spot, Region 3 is the kinematic nucleus, and Region 4 is the H$_2$ emission peak.}
\end{figure*}

 Figure~\ref{fig:spec} displays the rest-frame K band spectra, integrated over the regions shown on the H$_2$ map (Figure~\ref{fig:kband}).  The main emission lines are highlighted.  Specifically, we see many H$_2$ transitions as well as Br$\gamma$ and HeI.  The highest energy H$_2$ transition detected is the 3-2 S(3) transition. This is located at 2.2008 $\mu$m and only marginally visible in Region 8 of Figure~\ref{fig:spec}, although we have 2$\sigma$ detections of it in Regions 1, 2, and 3. This is an important detection since this transition can only be excited by high energy photons, and therefore points towards fluorescent excitation.  In Table~\ref{tab:flux} we present the integrated fluxes and fluxes in each region of all detected H$_2$ lines with error bars.  These error bars were derived by calculating the flux for a range of different continuum baselines and finding the standard deviation of these values. We also attempted to extract H$_2$ line fluxes from transitions in 
the H band, yet the stellar absorption features overpowered the emission lines and we were unable to get reliable flux measurements.    

Using the SINFONI data cubes, we can extract both a gas and stellar velocity map.  Figure~\ref{fig:vel} displays the stellar velocity field with the H$_2$ molecular gas velocity field in contours.  The stellar velocity field was calculated by fitting the three stellar CO photospheric absorption features.  Fitting with a set of 6 template stars (K3V, M0III, M0V, M4V, M5III and M5II), we used the \citet{2004PASP..116..138C} Penalised Pixel Fitting (PPXF) package.  All the template stars were observed with SINFONI on the same settings as the NGC 253 observations.  In addition, the pixels were binned using a Voronoi tessellation \citep{2003MNRAS.342..345C}, with a signal-to-noise of 50 per bin.  These binned spectra were then fit with the best combination of the 6 template stellar spectra varying velocity and dispersion.  Then, the systemic velocity of the galaxy, 243 km/s, was subtracted from the velocity field.  
\begin{table*}
\caption{Integrated fluxes for selected regions and over the full map of all detected H$_2$ transitions.  For the regions the units of flux are 10$^{-15}$ erg s$^{-1}$cm$^{-2}$.  Only detections above 3$\sigma$ are reported with the exception of 1-0 Q(4), where we report all fluxes over 2$\sigma$.  The errors on the integrated fluxes represent 1 standard deviation.  Each region is 12$\times$12 pixels, or 1.5$\times$1.5 arcseconds.}
 \begin{tabular}{|l||c|c|c|c|c|c|c|c|c|c|c|}
\hline

& 1-0 S(0)&1-0 S(1)&1-0 S(2)&1-0 S(3)&2-1 S(1)&2-1 S(3)&3-2 S(3)&1-0 Q(1)&1-0 Q(2)&1-0 Q(3)&1-0 (Q4)\\ 

\hline
Region 1 &0.7$\pm$0.03&2.8$\pm$0.07&1.0$\pm$0.08&2.6$\pm$0.3&0.5$\pm$0.1&0.2$\pm$0.04&0.1$\pm$0.01&2.8$\pm$0.3&0.7$\pm$0.1&2.6$\pm$0.4&-- \\
Region 2 &2.8$\pm$0.1&10.2$\pm$0.3&4.9$\pm$0.9&5.3$\pm$1.2&1.3$\pm$0.4&1.0$\pm$0.05&0.4$\pm$0.1&10.7$\pm$1.2&5.3$\pm$1.1&11.2$\pm$1.3&3.4$\pm$1.3\\
Region 3 &2.6$\pm$0.2&8.3$\pm$0.2&4.3$\pm$0.9&4.0$\pm$1.1&1.4$\pm$0.1&0.9$\pm$0.04&0.4$\pm$0.1&9.3$\pm$1.0&4.5$\pm$1.2&10.2$\pm$1.3&2.6$\pm$0.9\\
Region 4 &3.8$\pm$0.2&12.3$\pm$0.2&6.0$\pm$0.8&7.5$\pm$0.9&2.0$\pm$0.09&0.9$\pm$0.05&0.3$\pm$0.1&12.8$\pm$1.1&4.8$\pm$1.3&13.2$\pm$1.3&3.2$\pm$1.0\\
Region 5 &2.6$\pm$0.08&8.5$\pm$0.2&4.2$\pm$0.3&6.7$\pm$0.5&1.3$\pm$0.04&--&--&8.2$\pm$0.5&2.5$\pm$0.7&8.0$\pm$0.7&1.7$\pm$0.7 \\
Region 6 & 1.3$\pm$0.1&4.3$\pm$0.08&2.2$\pm$0.1&3.1$\pm$0.5&0.6$\pm$0.01&--&--&4.0$\pm$0.2&1.0$\pm$0.3&3.8$\pm$0.3&--\\
Region 7 &1.7$\pm$0.07&5.1$\pm$0.1&2.5$\pm$0.1&3.3$\pm$0.5&0.7$\pm$0.02&--&--&4.8$\pm$0.4&--&4.6$\pm$0.4&-- \\
Region 8 & 0.5$\pm$0.03&1.6$\pm$0.03&0.9$\pm$0.07&0.8$\pm$0.3&0.2$\pm$0.01&--&--&1.5$\pm$0.1&--&1.4$\pm$0.2&--\\
Region 9 & 0.3$\pm$0.009&0.8$\pm$0.009&0.4$\pm$0.02&0.5$\pm$0.1&0.07$\pm$0.003&0.03$\pm$0.003&--&0.8$\pm$0.08&0.4$\pm$0.05&0.7$\pm$0.06&0.3$\pm$0.1\\

Integrated&100$\pm$50&380$\pm$50&140$\pm$30&250$\pm$40&50$\pm$20&30$\pm$9&6$\pm$4&370$\pm$20&140$\pm$20&320$\pm$40&120$\pm$40\\
\hline
 \end{tabular}
\label{tab:flux}
\end{table*}

\section{The Kinematic Center of NGC 253}
\label{sec:kin}
The stellar velocity field and the H$_2$ velocity field show many general similarities, especially in the central region.  Specifically, both the stellar and gas velocity fields show an 'S'-shaped morphology in the velocity contours in the 0 km/s range.  This morphology is typical of galaxies with a bar potential, and supports the claim of a bar potential in NGC 253. 

\begin{figure*}
\centering
\includegraphics[width=8cm]{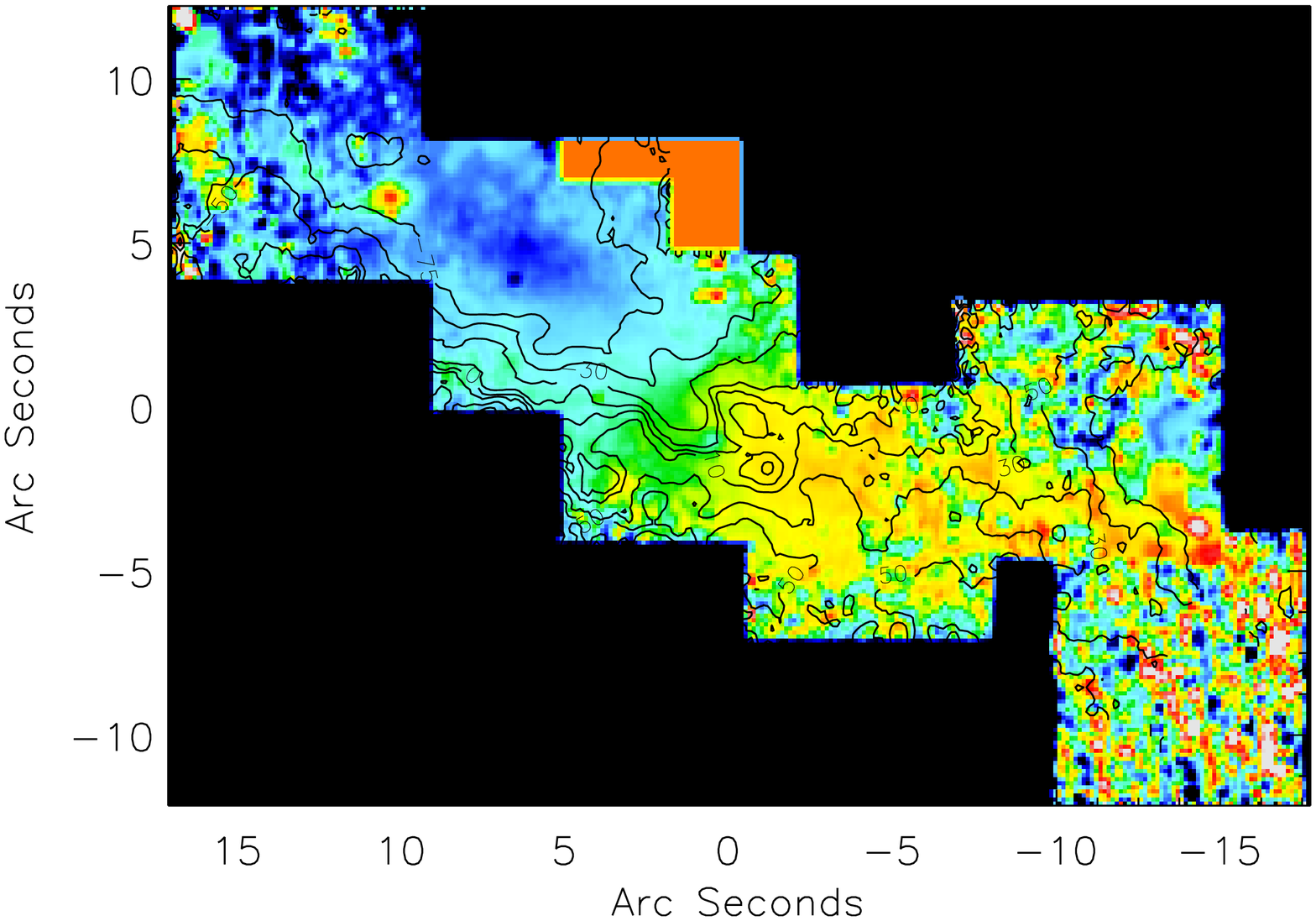}
\includegraphics[width=8cm]{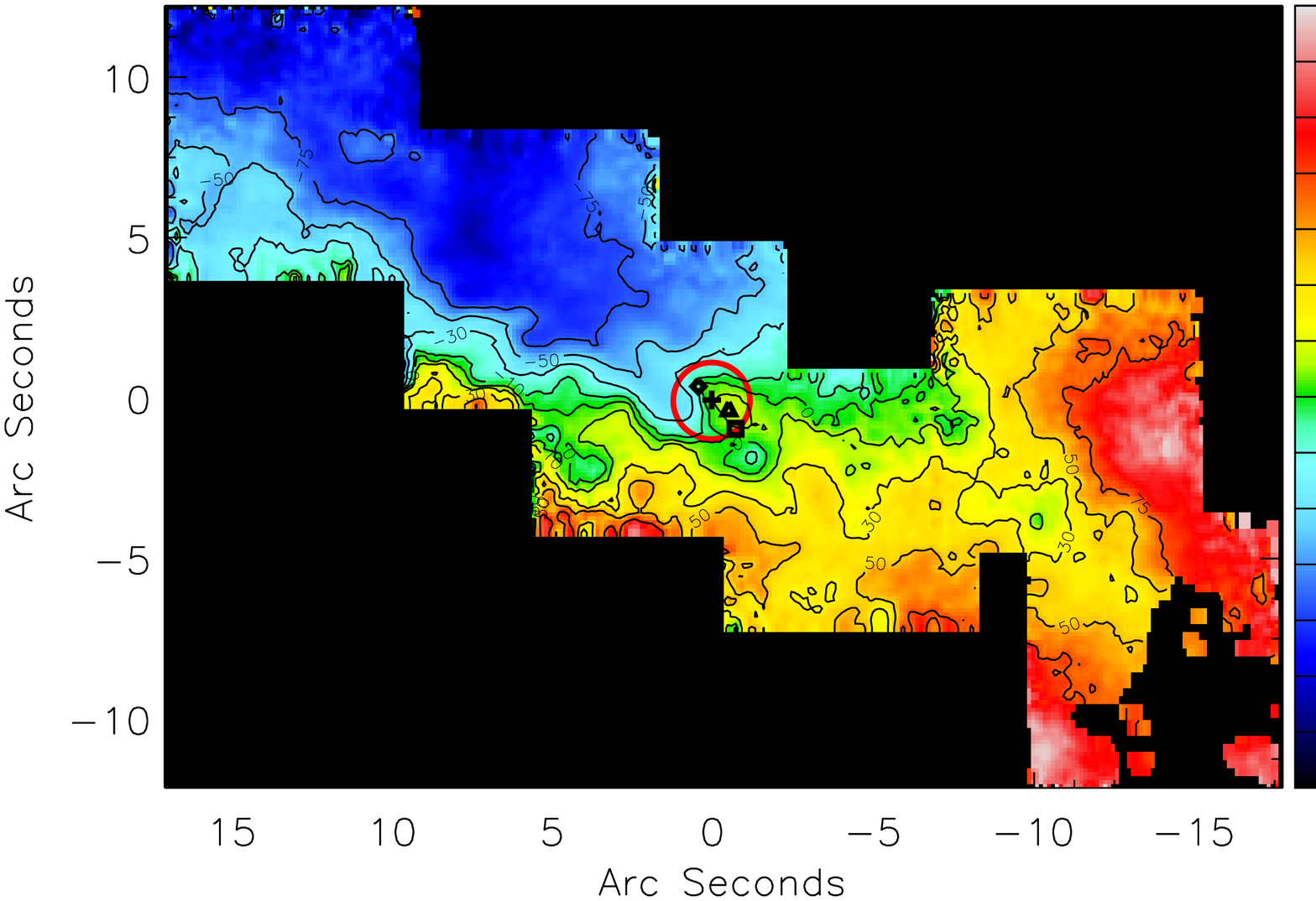}
\caption{Stellar velocity field (left) with H$_2$ 1-0 S(1) velocity contours and H$_2$ 1-0 S(1) velocity map with contours (right).  The contour levels are -75 ,-50 ,-30 ,-10 ,0 ,10 ,30 ,50 ,75 km/s.  The color bar describes the H$_2$ and stellar velocity in km/s. On the right panel, the red circle represents the 3$\sigma$ uncertainty for the center defined by \citet{2010ApJ...716.1166M}.  The plus sign is the center of this circle, and the kinematic center.  The diamond is TH2, the radio peak.  The triangle is X-1, the X-ray peak. The square is the point of highest symmetry.}
\label{fig:vel}
\end{figure*}

Since the H$_2$ velocity field is much cleaner than the stellar velocity field and they share a similar morphology, we will use the H$_2$ velocity to estimate the kinematic center of NGC 253.  In order to estimate this position, we create a mirrored reflection around a set of central points that could be the center.  By subtracting this mirrored map from the original map, we can find which pixel provides the most symmetric profile.  Using this method, we can determine the center pixel, or the pixel with the most symmetry.  To determine the absolute coordinates of this pixel, we employ the 2MASS K band image of NGC 253.  We anchor the SINFONI observations to the 2MASS observations by finding the position of the K-band peak.  We then use the offset of the kinematic center to the K band center to determine the absolute coordinates, $(\alpha,\delta)_{2000}$=(00$^h$47$^m$33$^s$.084,-25$^\circ$17'18''.42). 
\citet{2010ApJ...716.1166M} find the kinematic center at $(\alpha,\delta)_{2000}$=(00$^h$47$^m$33$^s$.14,-25$^\circ$17'17''.52) with a mean 3$\sigma$ error of r=1.2''.  Thus, our center is within this error and in good agreement with their result.  The difference between the two results is due to the method.  Muller-Sanchez used an inclined disk model to find their kinematic center, however the symmetry argument focuses on finding the point of highest symmetry.  This method is compromised due to two symmetric circular regions in the velocity field, the top circle is located within the red error circle and the symmetric counterpart is just below it and appears green.  Our method finds the center to be just between these two circles, at the radius of the Muller-Sanchez error radius, represented as a square.
Due to this symmetric point, we consider the Muller-Sanchez center to be the true kinematic center, which is also clear from the inflection point of the bar profile 'S' shape seen in the H2 velocity field.  Since the H2 inflection point coincides perfectly with the Muller-Sanchez center, it proves that there is no true offest between the H2 and CO measured centers.We therefore independently confirm the kinematic center proposed by \citet{2010ApJ...716.1166M}.  In Figure~\ref{fig:vel}, the H$_2$ 1-0 S(1) velocity map is presented with the \citet{2010ApJ...716.1166M} 3$\sigma$ center circled in red and their proposed kinematic center overlaid with a white plus sign.  For completeness, the radio peak TH2 and the x-ray peak X-1 are both overplotted as a diamond and triangle respectively.    

\section{Gas Excitation Mechanisms in NGC 253}
\label{sec:h2}
In general, the morphologies of the K band continuum, H$_2$, Br$\gamma$, and [FeII]$_{1.64}$ emission are dissimilar.  None of the peaks are coincident with each other with the exception of the K band and Br$\gamma$ peaks at the super star cluster.  There are certain regions where the H$_2$ matches the Br$\gamma$, while some H$_2$ clumps match the [FeII]$_{1.64}$ emission. This suggests that the H$_2$, Br$\gamma$ and [FeII] are being excited through different physical processes.   

The Br$\gamma$ peak, tracing massive, young star formation, is coincident with the K band continuum peak, at the location of the super star cluster.  In general, the K band continuum traces the older medium and low mass populations of stars in contrast to the youngest, most massive stars powering the Br$\gamma$ emission.  There is extended emission in the northeastern direction of the Br$\gamma$ peak, but no other strong emission regions.   

[FeII] is another strong NIR diagnostic line, which traces shocks.  In interstellar space, iron atoms are typically fully locked into dust grains. Shock fronts associated with supernova remnants (SNRs) may cause very efficient grain destruction through thermal sputtering \citep{1994ApJ...433..797J}.  This releases the iron into the gas-phase where it is singly ionized by the interstellar radiation field \citep{2000ApJ...528..186M}.  In the extended post shock region, Fe$^+$ is excited by electron collisions, rendering it a strong diagnostic shock tracer.  The [FeII] peak is located slightly northeast of the H$_2$ peak (Region 4).  It shows diffuse emission consistently along the disk, but does not show any peak at Region 2, the location of the super star cluster.  From \citet{2012A&A...540A.116R} we can calculate the supernova rate (SNrate) from the FeII luminosity.  In Table~\ref{tab:snr} the calculated SNrates for each region and the whole galaxy is given.  The small amount of [FeII] flux and 
correspondingly low supernova rate suggests that the super star cluster in Region 2 is a young cluster.  \citet{2009MNRAS.392L..16F} suggests that Region 2 is a starburst of 6 Myrs, which is also consistent with our age calculations.  Our integrated supernova rate in the inner 300 pc of the galaxy, is 0.2 yr$^{-1}$, which is well matched to previous supernova rate measurement of 0.1-0.3 \citep{1980ApJ...238...24R,1988ApJ...325..679R,1997ApJ...488..621U}.

Unlike Br$\gamma$ and [FeII], the mechanism which excites the H$_2$ gas is not as clear.  The NIR is rich with ro-vibrational H$_2$ lines, which can either be excited thermally, through shocks, or by fluorescence, through UV photons from O and B stars.  In the case of shock excited H$_2$ emission, the gas is thermalized and the energy levels are populated in a ``bottom-up'' manner.  The resulting gas temperature is around 1000 K and thus there is little H$_2$ emission from the $v\geq3$ states, which have temperatures above 15,000 K.  However, in the case of excitation by UV photons, the H$_2$ molecule absorbs a highly energetic photon and is excited to an upper electronic energy level and then proceeds to cascade downwards.  This is considered populating the energy levels from the ``top-down'', which results in exciting higher H$_2$ energy levels.  In this way, we can use the ratio of certain H$_2$ lines to discriminate which excitation mechanism is exciting the H$_2$ emission.  However, the case of UV 
excitation is sometimes degenerate.  In the case of high density, the molecules can collide frequently enough to become thermalized in the lower energy levels, which results in line ratios equivalent to the case of shock excitation. For the lines to be thermalized, the density in the region must be comparable to the critical density of the lines, $1.7\times10^5$ cm$^{-2}$ for 1-0 S(1) and $1.2\times10^5$ cm$^{-2}$ for 2-1 S(1) \citep{1989ApJ...338..197S}.  

In order to determine which excitation mechanisms are dominating which regions, we can use the information from the Br$\gamma$ and [FeII] tracers.  Specifically, if the H$_2$ is undergoing excitation by UV photons in a dense environment, this region should also be bright in Br$\gamma$ emission if the excitation is by O stars.  Conversely, if the H$_2$ is being excited by shocks, then the [FeII] emission might also be bright in this region.   

In Region 2, coincident with the Br$\gamma$ peak, UV photon excitation is the dominant mechanism and the density of this region is accounting for the lower excitation levels being populated leading to the lower H$_2$ line ratio.  In Region 6, there is little Br$\gamma$ emission.  However, the [FeII] emission shows a bright clump.  Therefore, in this region the emission lines suggest that the H$_2$ is being excited by shocks.  However, in the other regions, there is both diffuse [FeII] and diffuse Br$\gamma$.  In this case, we cannot determine the dominant mechanism by only studying the Br$\gamma$ and [FeII] emission. Since H$_2$ can also be excited by slow shocks, too slow to destroy dust grains, or by B stars, which are not energetic enough to excite Br$\gamma$, we must use other diagnostics to probe the excitation mechanisms.

\begin{table}
\caption{Calculated supernova rates for each region and the integrated galaxy based on [FeII] luminosity.  The integrated value is integrated over the full map.  Supernova rates were calculated with the relation between [FeII] luminosity and SNrate from \citet{2012A&A...540A.116R}.}
 \begin{tabular}{|l||c|c|}
\hline
Region&   [FeII] Luminosity & SNrate \\
      & 10$^{37}$ erg s$^{-1}$ &10$^{-3}$ yr$^{-1}$ \\
\hline
Region 1& 0.3  &  0.04    \\
Region 2& 0.3  & 0.04    \\
Region 3& 9.8  &1.6    \\
Region 4& 11.3  &1.8    \\
Region 5& 17.1  &2.8      \\
Region 6& 4.1  &0.7      \\
Region 7& 0.3  &0.04     \\
Region 8& 0.5  &0.08      \\
Region 9& 0.4  &0.06          \\
Integrated& 140.9 &23.5       \\
\hline
 \end{tabular}
\label{tab:snr}
\end{table}

One diagnostic is the 2-1 S(1) to 1-0 S(1) transitions at 2.2478 $\mu$m and 2.1218 $\mu$m respectively.  This line ratio is commonly used because both lines are relatively bright and have similar wavelengths, so they can usually be taken in the same spectrum and suffer identical extinctions.  Predictions for the ratio of these two lines due to excitations by shocks range from 0.1 to 0.2 \citep{1978ApJ...220..525S}, while for photo-excitation it is predicted to be 0.53-0.56 \citep{1987ApJ...322..412B}. 

\subsection{Shocks vs Fluorescence}

\begin{figure*}
\centering
\includegraphics[trim=0cm 0.1cm 0cm 0cm, clip=true,width =17cm]{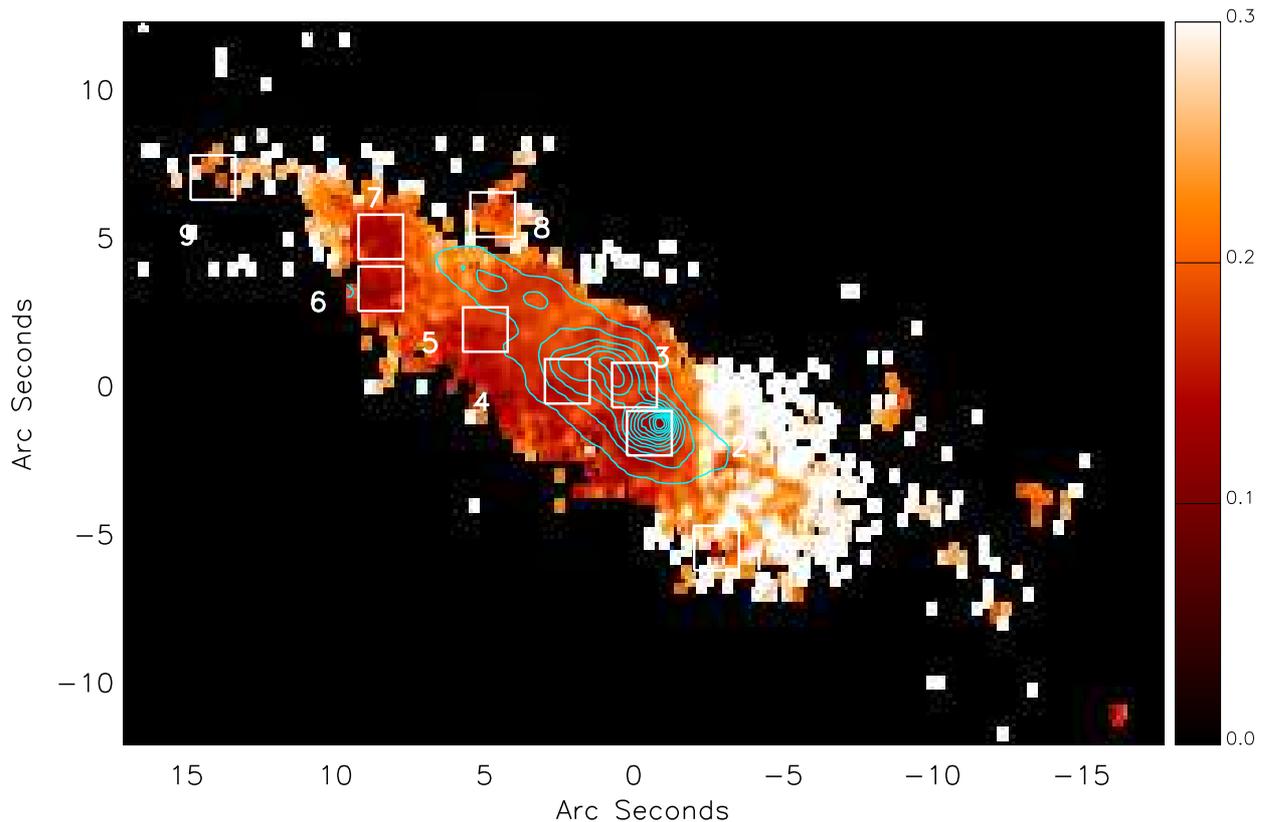}
\caption{The ratio of 2-1 S(1)/1-0 S(1) H$_2$ lines with the Br$\gamma$ line map over plotted in contours.  The white rectangles are the same regions as defined in Figure~\ref{fig:kband}.  The color bar values represent the value of the ratio.  The offset from the kinematic center at position $(\alpha,\delta)_{2000}=(00^{h}47^{m}33^{s}.14,-25^{\circ}17'17''.52)$ is given on each axis.  Ratio values where the line flux of either transition is below 10\% of the maximum flux of that line are masked.}
 \label{fig:h2ratio}
\end{figure*}

In Figure~\ref{fig:h2ratio}, we present a map of the 2-1 S(1)/1-0 S(1) ratio with the contours of the Br$\gamma$ line map and the rectangular regions from Figure~\ref{fig:kband} overlaid.  The 2-1 S(1) line is significantly weaker than the 1-0 S(1) transition thus this image has less signal-to-noise.  We have therefore applied a mask to the map, filtering out values lower than 10\% of the maximum flux for both line transitions.  The values of the 2-1 S(1)/1-0 S(1) ratio range continuously from 0 to 0.5, showing that we have both thermalized and fluorescently excited gas in the galaxy.  Both the mean H$_2$ ratio for each region and the ratio of the integratred 2-1 S(1) and 1-0 S(1) line fluxes is listed in Table~\ref{tab:h2lineratio}.  The discrepency in these two values is due to the fact that in the left column, we find the mean value of the H$_2$ ratio at each pixel for each region, whereas in the right column we first integrate the flux in each region for the two lines, and then divide 
the fluxes.  The latter method is more sensitive to high surface brightness quantities, and shocks excited material has higher surface brightness than that of fluorescent excitation \citep{1978ApJ...220..525S}.  Therefore, the values in the right column are weighted towards shock excitation, and thus show lower ratios for each region.  Averaging the ratio on a pixel-by-pixel basis, as done in the left column, does not present such a bias.  In regions with very high ratios ($\geq 0.3$) we can confidently conclude that the molecular hydrogen is being excited by UV photons from O and B stars.  Similarly, there are a few regions where the ratio is distinctly lower than the average value of $\sim 0.2$.  These regions are region 2, 6, and 7 from Figure~\ref{fig:h2ratio}.  
It is interesting to note that these bright regions all correlate with an H$_2$ clump or luminosity peak.  In these regions we cannot 
determine whether the low H$_2$ line ratio is indicating excitation by shocks, or a very dense clump being excited by UV photons.  However, we can place an upper limit on the amount of shocked gas.  Assuming that all gas with a 2-1 S(1)/1-0 S(1) ratio lower than 0.2 is shock excited, we see that 48\% of the total 1-0 S(1) H$_2$ flux may be from ``shocked'' gas.  However, this percentage is including all regions with a ratio lower than 0.2, including regions of high density where the gas is fluorescently excited but thermalized.  We can at least partially separate this effect by filtering out the pixels with the highest Br$\gamma$ flux ($\geq 1\times10^{-16}$ erg s$^{-1}$ cm$^{-2}$), since these regions are dense regions of known fluorescent excitation.  As seen in the Br$\gamma$ contours in Figure~\ref{fig:h2ratio}, this affects Regions 2, 3, and 4.  Excluding these regions and any pixel with high Br$\gamma$ flux, no more than 29\% of all H$_2$ is shock excited, and the actual amount 
may be less.

\begin{table}
\caption{Mean H$_2$ 2-1 S(1)/1-0 S(1) line ratios in each region of Figure~\ref{fig:h2lines} (left column) and the ratio of the 2-1 S(1) and 1-0 S(1) fluxes (right column).  The integrated value is integrated over the whole map.}
 \begin{tabular}{|l||c|c|}
\hline
Region& $\overline{\left ( \frac{2-1 S(1))}{1-0 S(0))} \right )}$&$\frac{\sum 2-1 S(1)}{\sum 1-0 S(0)}$  \\
\hline
Region 1&   0.26$\pm$0.01&0.17$\pm$0.04  \\
Region 2&   0.14$\pm$0.02&0.13$\pm$0.04   \\
Region 3&   0.17$\pm$0.03&0.17$\pm$0.01      \\
Region 4&   0.16$\pm$0.02&0.16$\pm$0.01    \\
Region 5&   0.17$\pm$0.02&0.15$\pm$0.01     \\
Region 6&   0.15$\pm$0.02&0.14$\pm$0.003     \\
Region 7&   0.15$\pm$0.04&0.15$\pm$0.01     \\
Region 8&   0.21$\pm$0.02&0.16$\pm$0.003      \\
Region 9&   0.23$\pm$0.01&0.08$\pm$0.01    \\
Integrated& 0.27$\pm$0.06&0.13$\pm$0.06      \\
\hline
 \end{tabular}
\label{tab:h2lineratio}
\end{table}

\subsection{Tracers of Fluorescence}
If strong shocks dominate the excitation of H$_2$ in NGC 253, we would expect to see a good correlation between H$_2$ and [FeII], which we do not.  If fluorescence is the dominant mechanism, we would expect the Br$\gamma$ morphology to share more similarities with that of the H$_2$.  We also see no correlation between H$_2$ and Br$\gamma$ emission.  However, Br$\gamma$ only traces the most massive and young stars, predominately the O stars.  In order to excite Br$\gamma$, the star must emitting Lyman continuum photons ($\lambda < 912 \AA$, 13.6 eV), yet H$_2$ can be excited by lower energy photons as well ($\lambda < 1100 \AA$, 11.2 eV). 

Similarly to H$_2$, polycyclic aromatic hydrocarbons (PAHs) have a lower excitation energy than HI ($\lambda \sim 4200 \AA$ for N$_C>50$ where N$_C>50$ is the number of carbon atoms; \cite{1994ApJ...427..822B}) and thus are excited not only by O stars but also by slightly less massive, B-type stars.  Therefore, the PAH emission traces the more general fluorescently excited population, including excitation by both O and B stars. By comparing the morphology of the H$_2$ to that of PAH emission, we can determine if it is indeed the B stars that are predominately responsible for exciting the H$_2$ regions, and why neither the Br$\gamma$, [FeII] or K band continuum matches its morphology.  

\begin{figure*}
\centering
\includegraphics[trim=0cm 1.2cm 3cm 0cm, clip=true,width=6.cm]{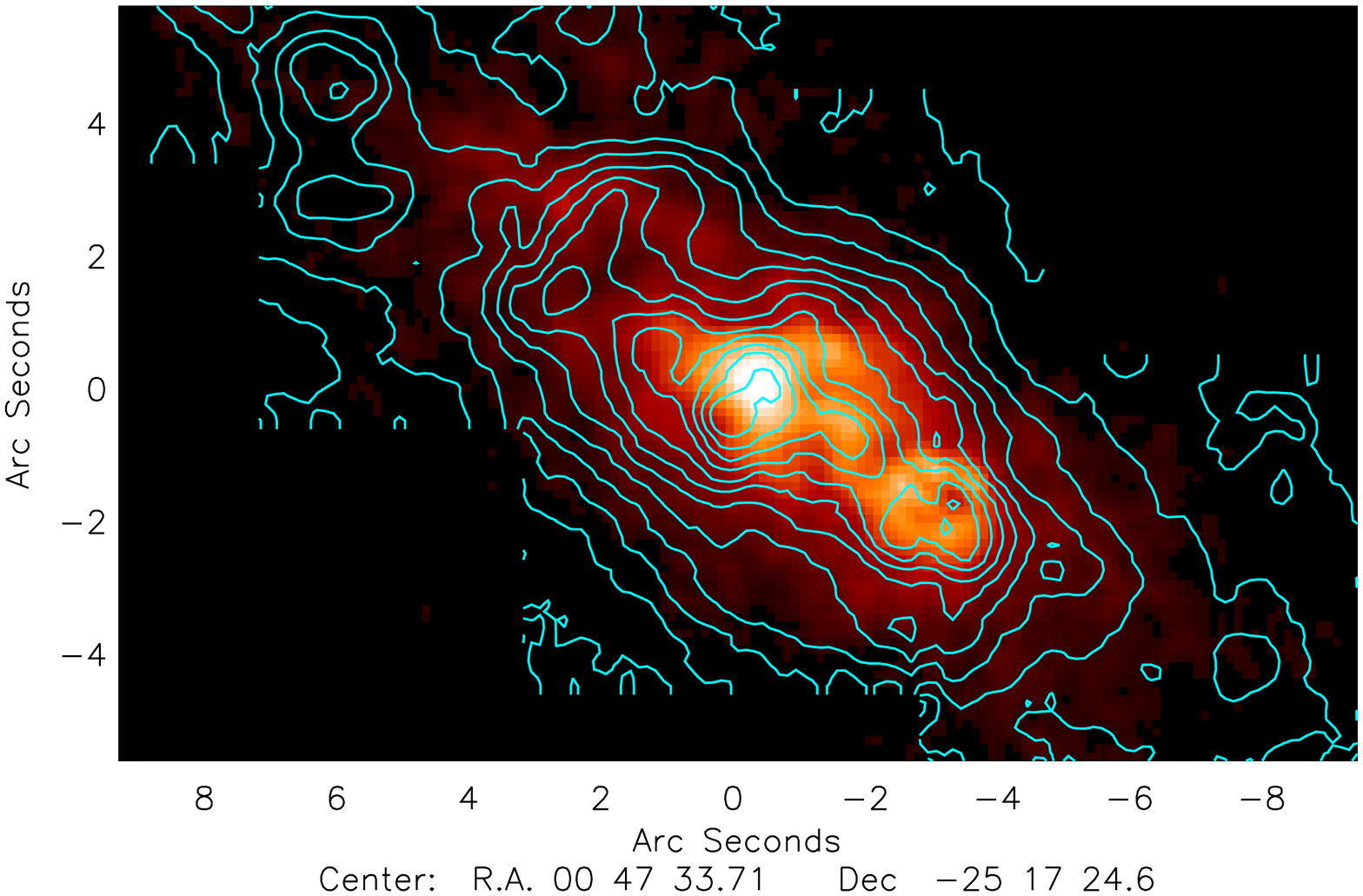}
\includegraphics[trim=0cm 1.2cm 3cm 0cm, clip=true,width=6.cm]{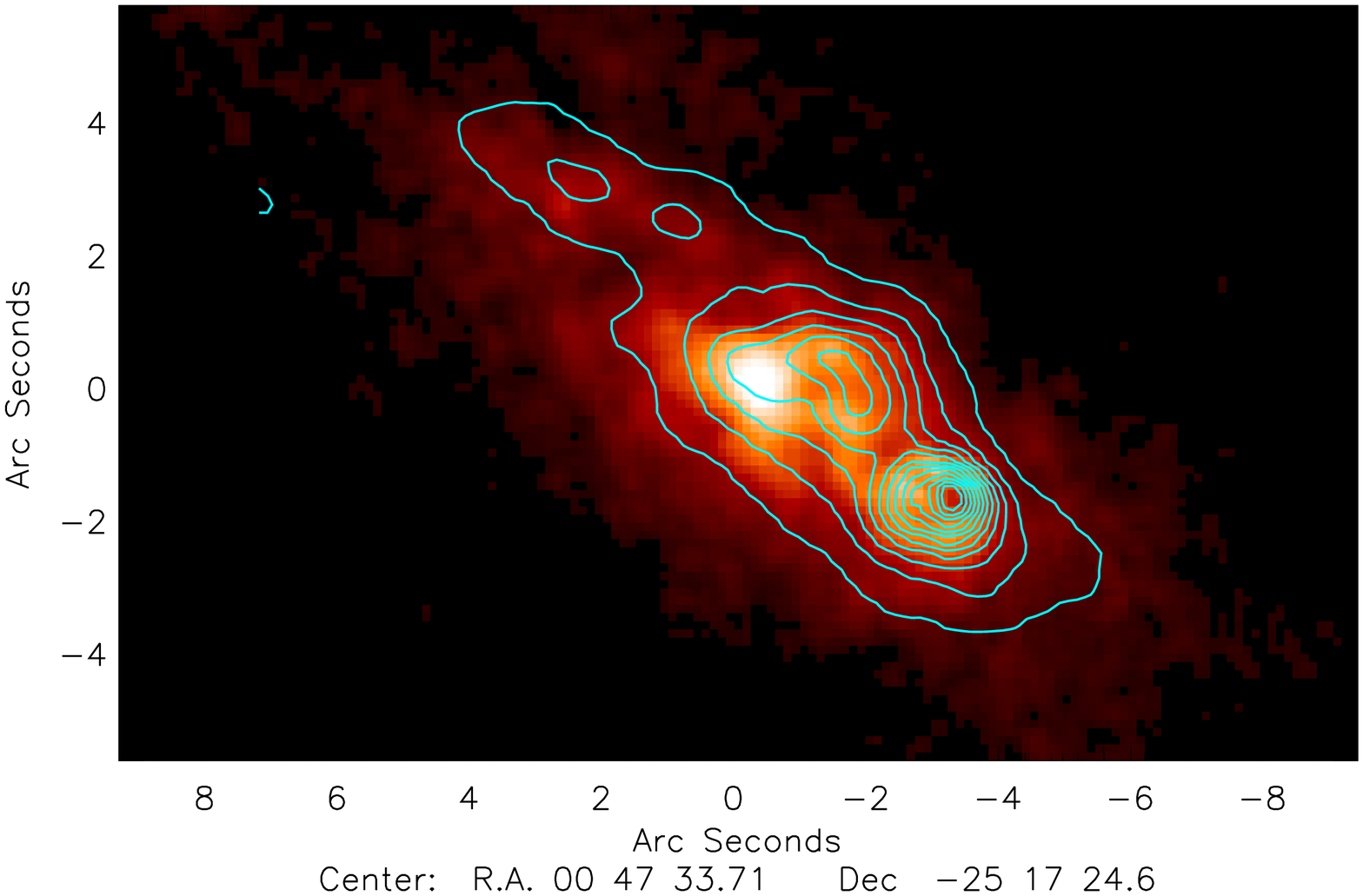}
\includegraphics[trim=0cm 1.2cm 3cm 0cm, clip=true,width=6.cm]{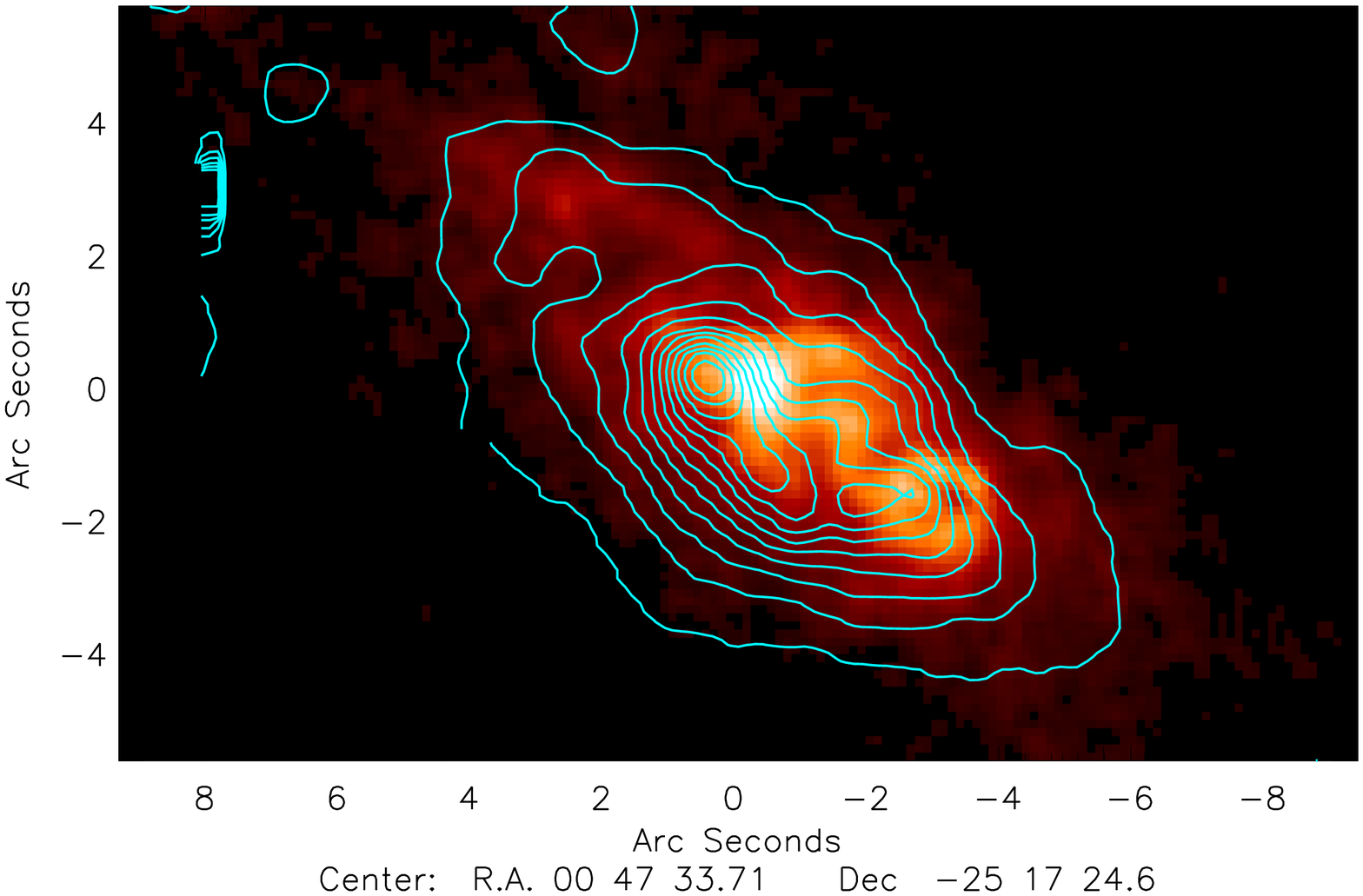}
\includegraphics[width=17cm]{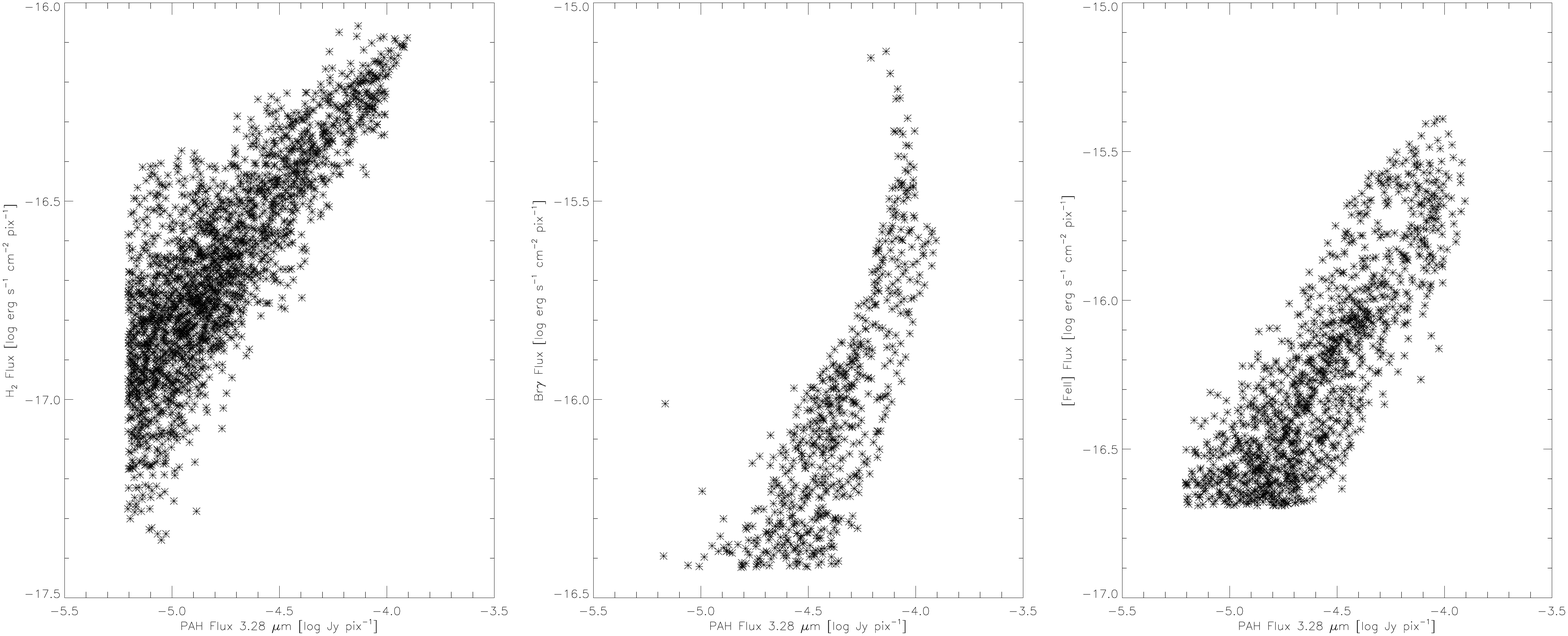}
\caption{Top Panels: PAH 3.28 $\mu$m continuum subtracted image from ISAAC.  Left panel has H$_2$ 1-0 S(1) contours, middle panel has Br$\gamma$ contours, right panel has [FeII]$_{1.64}$ contours.  The offset from the kinematic center at position $(\alpha,\delta)_{2000}=(00^{h}47^{m}33^{s}.14,-25^{\circ}17'17''.52)$ is given on each axis.  The kinematic center is from \citet{2010ApJ...716.1166M}.  Bottom Panels: Pixel-pixel plots of PAH flux vs H$_2$ flux (left), Br$\gamma$ flux (center), and [FeII] flux (right).  The pixels of the ISAAC PAH observations were rebinned to match those of the SINFONI observations.}
\label{fig:pah}
\end{figure*}

In Figure~\ref{fig:pah}, we present an ISAAC continuum subtracted PAH map at 3.21$\mu$m.  The left panel is overlaid with H$_2$ 1-0 S(1) contours, the middle panel is overlaid with Br$\gamma$ contours, and the right panel is overlaid with [FeII]$_{1.64}$ contours. The two maps were centred with respect to one another by matching the K band continuum peak of the SINFONI maps to the 3.21 continuum peak of the ISAAC continuum maps.

From these images, the morphology of the PAH emission matches the H$_2$ emission, better than both the Br$\gamma$ and [FeII].  The Br$\gamma$ peak is coincident with secondary PAH peak, meaning that in this region, there are many O stars producing photons $E > 13.6 eV$. However, there is little Br$\gamma$ emission coincident with the primary PAH peak, showing that these regions are mainly excited by B stars.  The [FeII] peak shows little similarity to the PAH emission.  However, the H$_2$ peaks are coincident with both the primary and secondary PAH peaks and shares a generally similar morphology in the diffuse PAH emission.  It is interesting to note that the elongated feature seen in the K Continuum, [FeII] and Br$\gamma$ maps in Figure~\ref{fig:bg}, is also present in the PAH map, but not in the H$_2$.  Since we still have a high amount of diffuse emission in this region, it is most likely that there is H$_2$ emission originating from this feature, but it is just relatively dimmer than the nearby 
H$_2$ peaks, which therefore overshadow it.    

In order to test this correlation in a more quantitative sense, we can perform a pixel-pixel analysis of the correlation of our line maps to that of the PAH 3.21 $\mu$m emission.  After re-binning the ISAAC PAH map to match the pixel size of SINFONI, we plot the flux of H$_2$, Br$\gamma$ and [FeII] as compared to the PAH flux.  Figure~\ref{fig:pah} depicts these correlation plots, which show how well H$_2$, Br$\gamma$ and [FeII] correlate with the PAH flux.  The Br$\gamma$ vs PAH plot shows an interesting morphology, in which the emission seems to be coming from two different areas.  First, there is an approximately linear correlation for lower Br$\gamma$ fluxes.  This correlation is most likely due to the diffuse Br$\gamma$ emission caused by the same photons exciting the PAHs.  However, there is a second, non-linear, nearly vertical component of extremely high Br$\gamma$ flux and medium strength PAH flux.  These points correspond to the location of the super star cluster, which is only the secondary 
flux peak for PAH, but the dominating source of Br$\gamma$ flux.  The [FeII] also seems to correlate with PAH flux with a large, but constant spread, suggesting that these tracers originate in the same general region but without a direct physical correlation, emphasized by the displacement of the [FeII] peak and the PAH peak.  The H$_2$ also shows a linear correlation with the PAH flux.  However, this correlation has a very large spread at low fluxes and gets continually tighter as both H$_2$ and PAH fluxes increase.  We can see this same trend in the morphologies as well.  There is a strong correlation, both spatially and numerically for the peaks of the H$_2$ and PAH linemaps, however in the diffuse regions the correlation no longer holds.


The strong, linear correlation between H$_2$ and PAH morphology in the central region of NGC 253 gives added confidence that most of the central emission in NGC 253 (Regions 2, 3, and 4) is fluorescently excited.  In addition, the correlation between the PAH morphology and H$_2$ morphology indicates that the H$_2$ in this area, is being excited by the same stars that are exciting the PAHs, mainly B-type stars.

\section{PDR and Shock Models}
\label{sec:mod}
Although the morphological correlation of H$_2$ to fluorescently excited PAHs is a strong indicator that this is the dominant mechanism in NGC 253, we can also compare the H$_2$ line ratios in each region to the predicted values from both shock and PDR models for a more quantitative analysis.  In Figure~\ref{fig:h2ratio}, the ratio is lower than predicted for pure UV excitation, although the correlation to the PAH maps indicates that UV fluorescence is the dominant mechanism.  In order to study this further, we present a diagnostic excitation level diagram.  This diagram can be seen in Figure~\ref{fig:h2lines} and plots the upper energy level ($E/k$) against the column density distribution($N$) divided by degeneracy (g) where:

\begin{equation}
 \frac{N_{obs}(v,J)}{g_J}=\frac{4\pi\lambda}{hc}\frac{I_{obs}(v,v',J,J')}{A(v,v',J,J')}
\end{equation}

Here, N$_{obs}(v,J)$ is the observed column density of a specific upper electronic energy level (v,J), g$_J$ is the degeneracy of this level, A(v,v',J,J') is the Einstein-A radiative transition probability from \citet{1998ApJS..115..293W}, and I$_{obs}$ is the observed line flux for each transition.  The column density is normalized by the 1-0 S(1) column density divided by its degeneracy ($g_J$).  In addition, we can plot the best fit Maxwell-Boltzmann distribution to the $v=1$ transitions.  The slope of this line tells us the excitation temperature in the region.  

\begin{figure*}
\centering
\includegraphics[width=5.5cm]{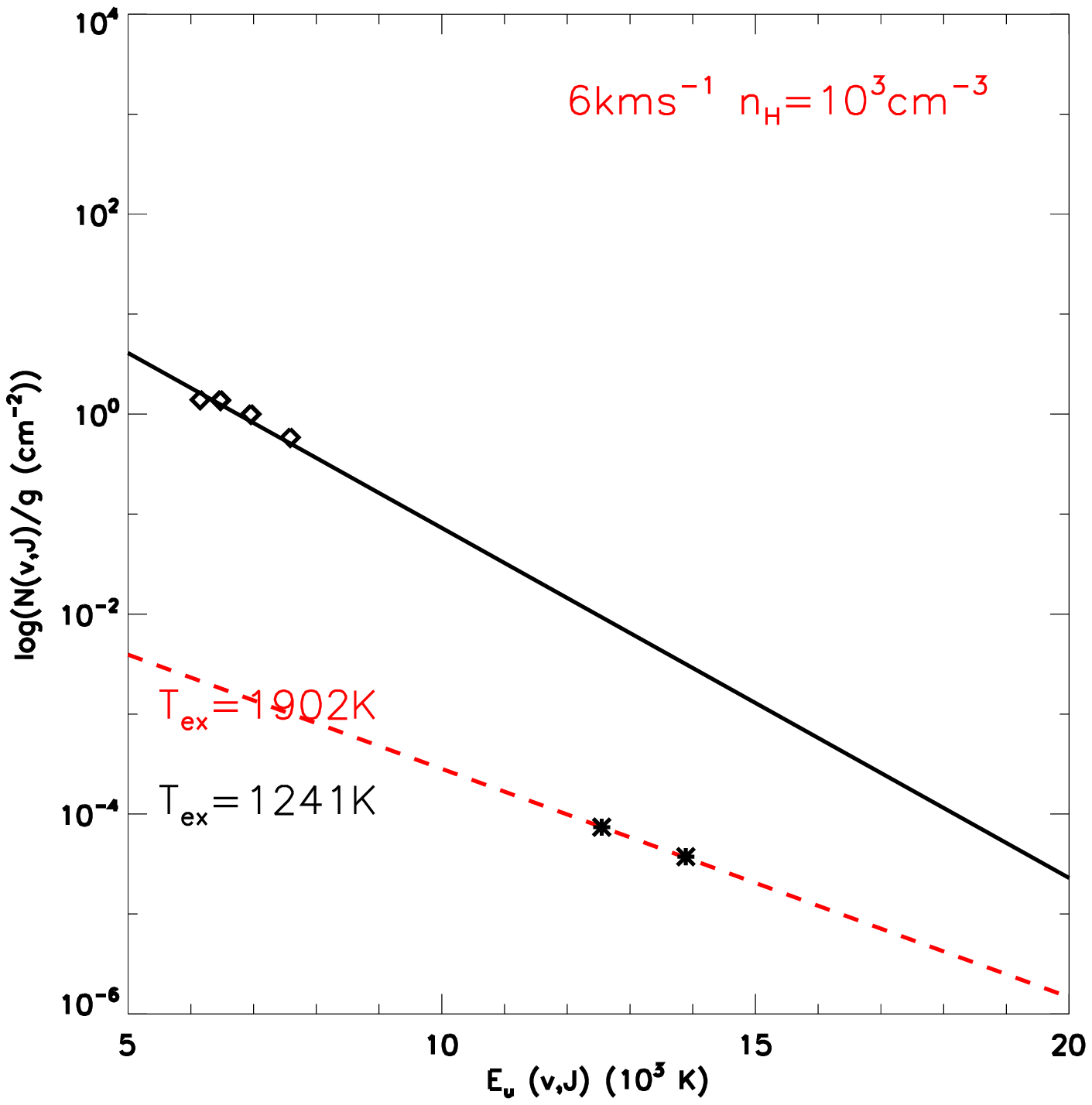}
\includegraphics[width=5.5cm]{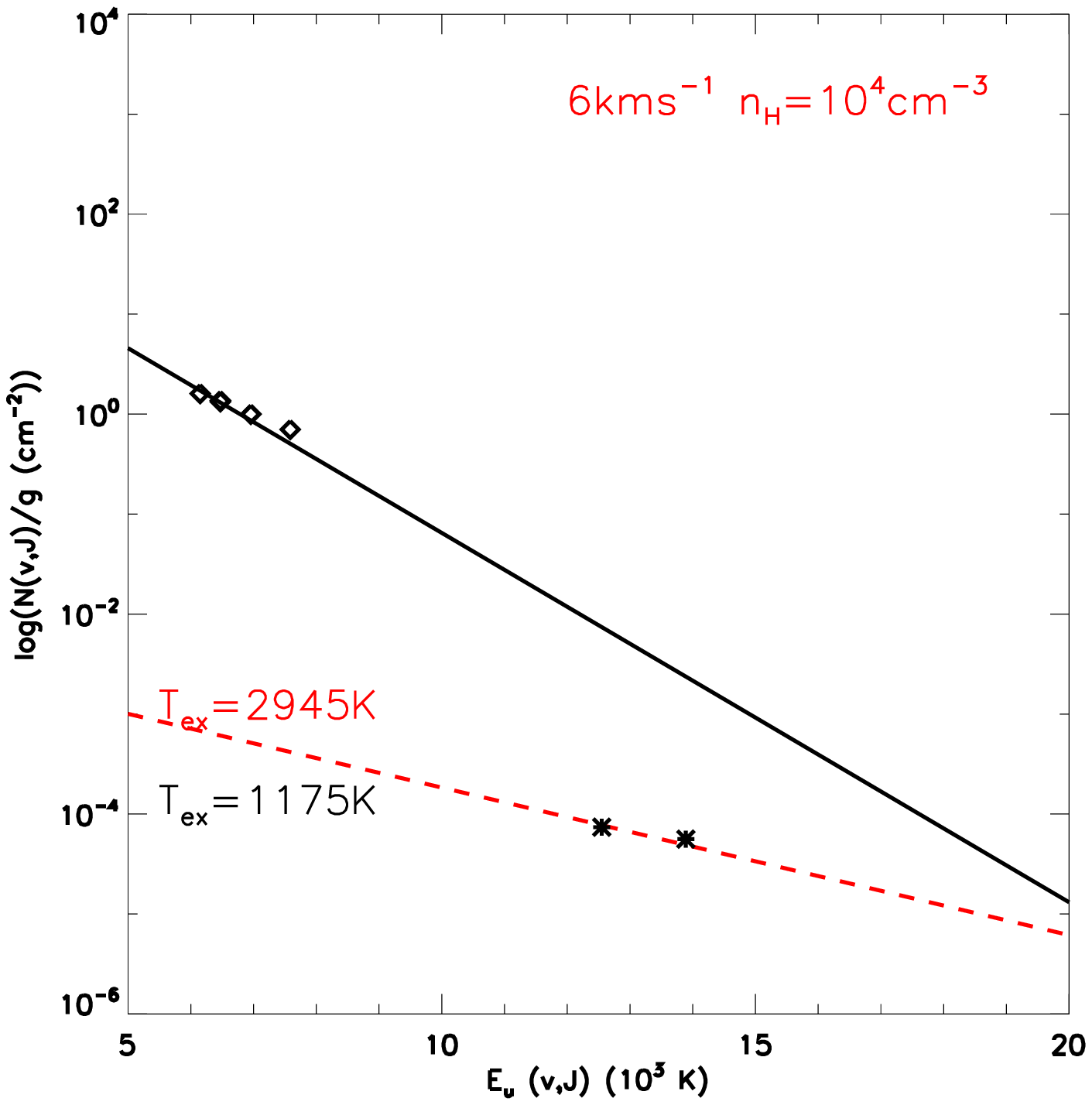}
\includegraphics[width=5.5cm]{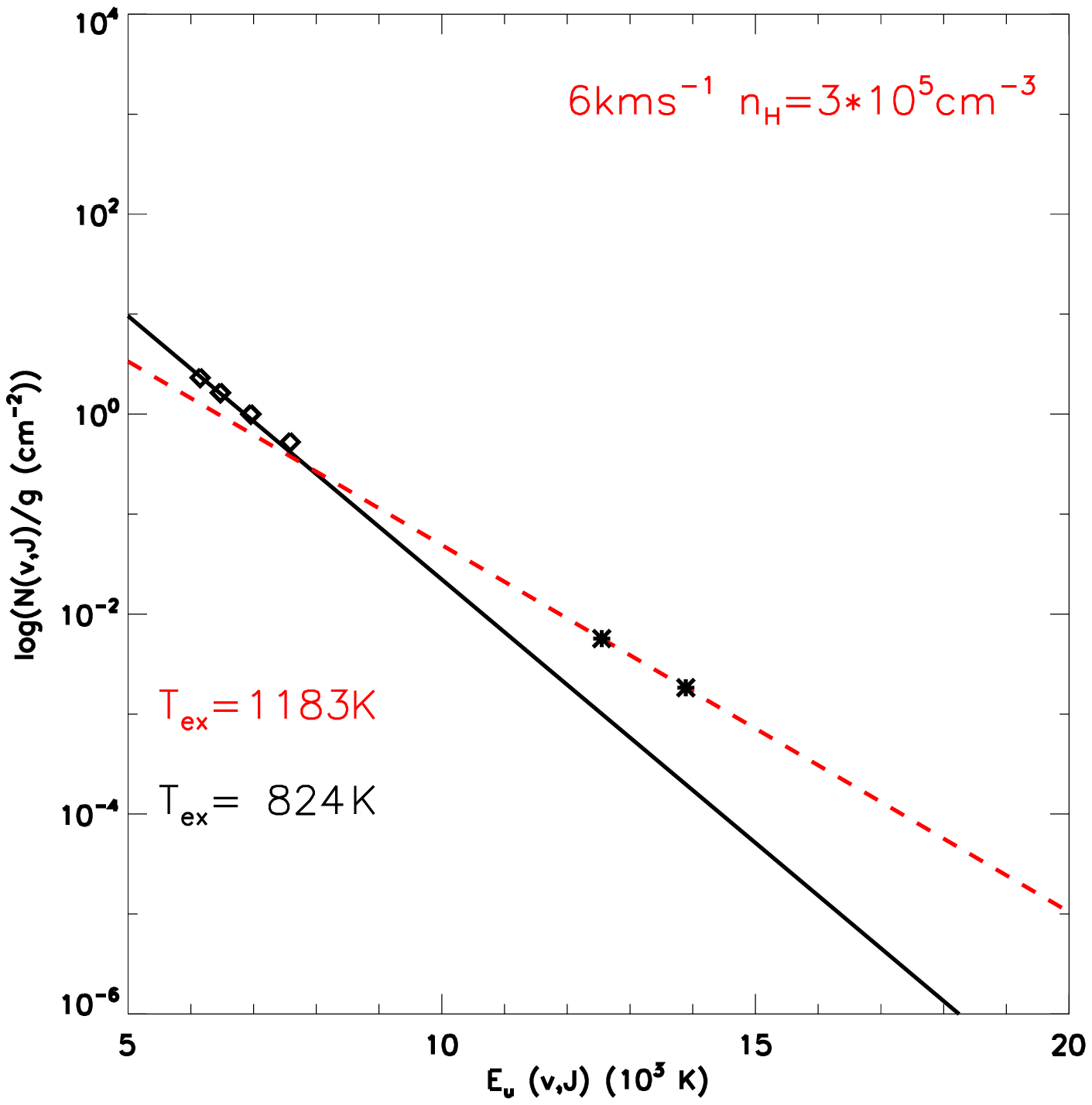}
\includegraphics[width=5.5cm]{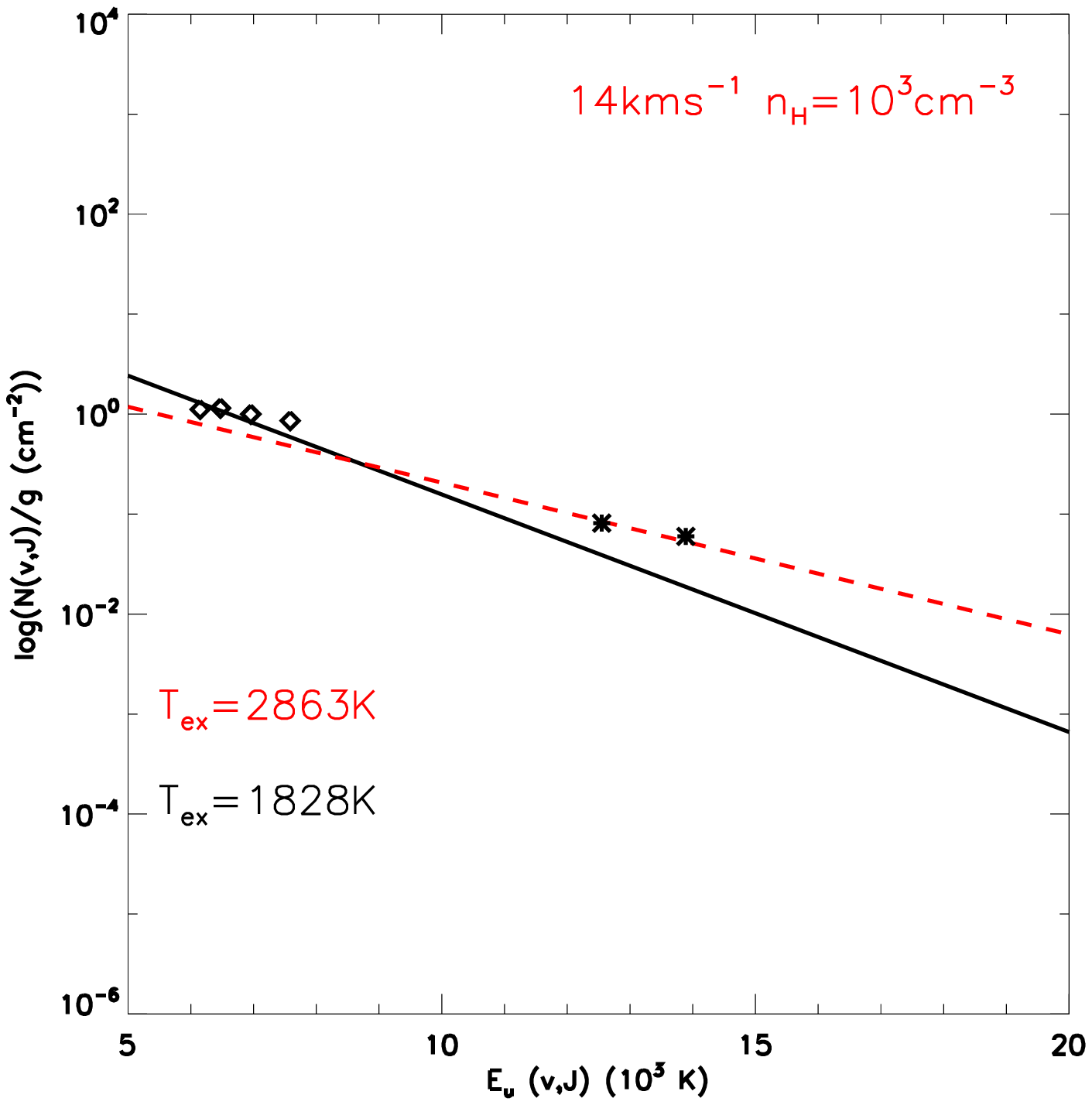}
\includegraphics[width=5.5cm]{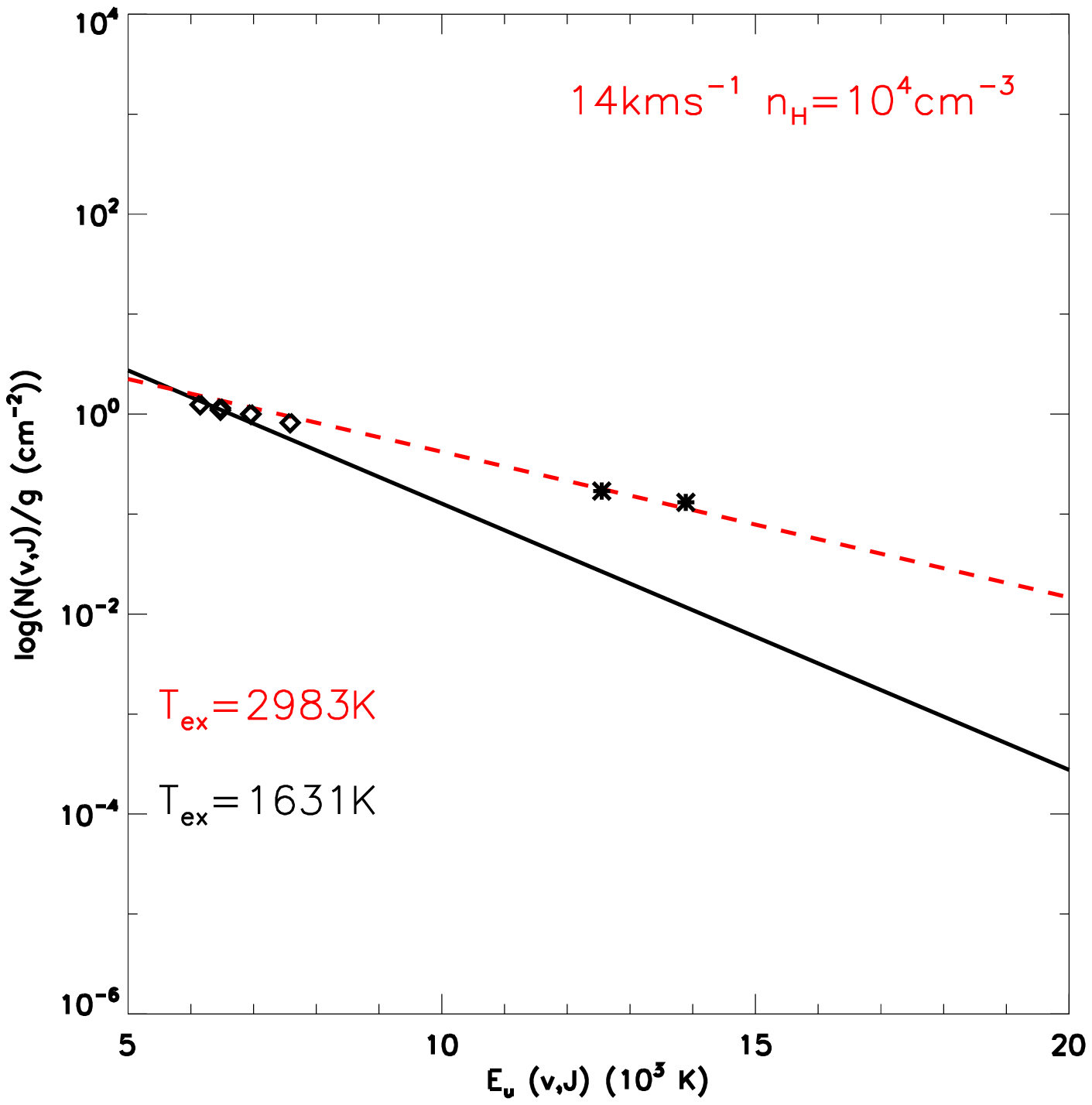}
\includegraphics[width=5.5cm]{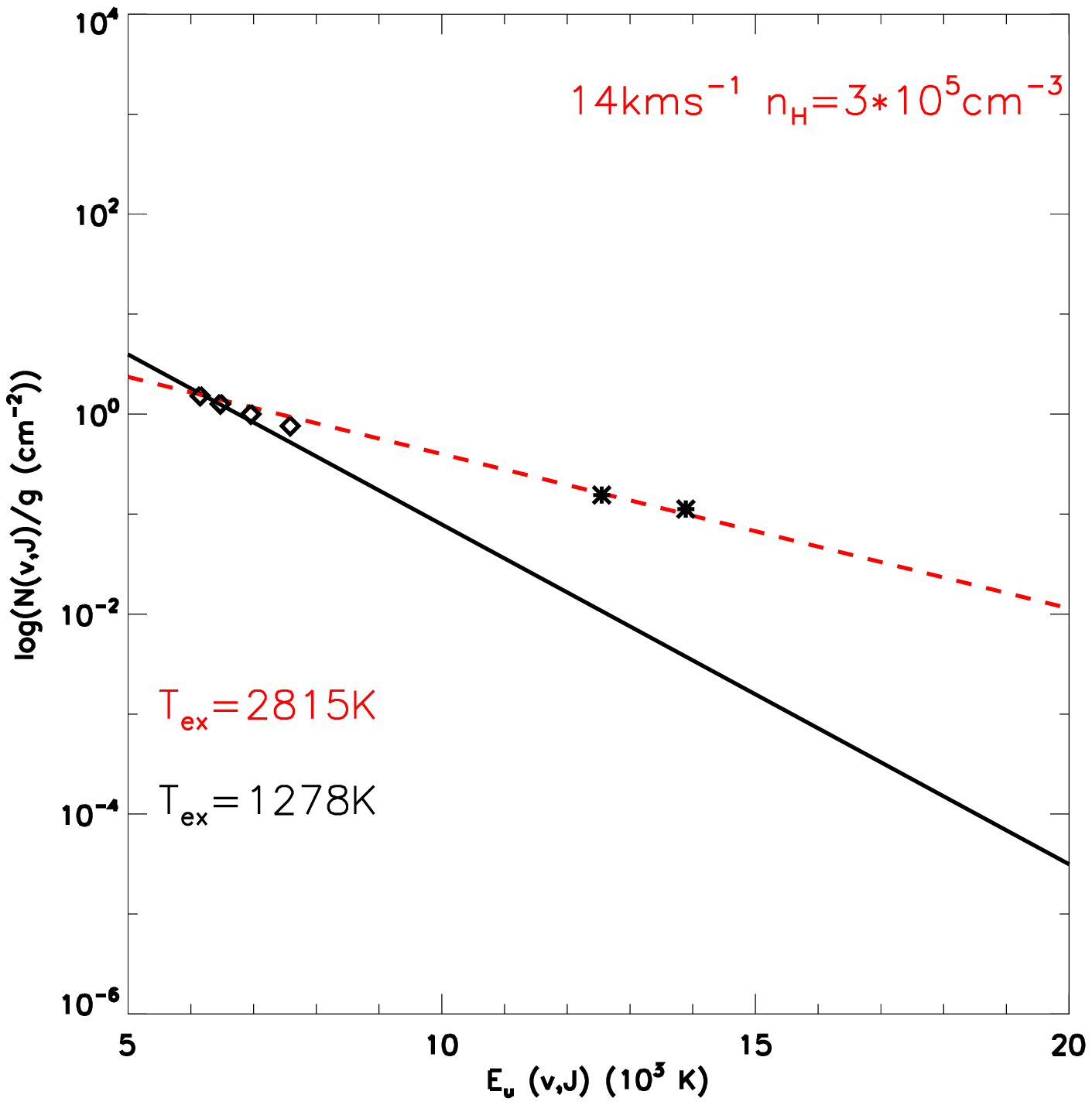}
\includegraphics[width=5.5cm]{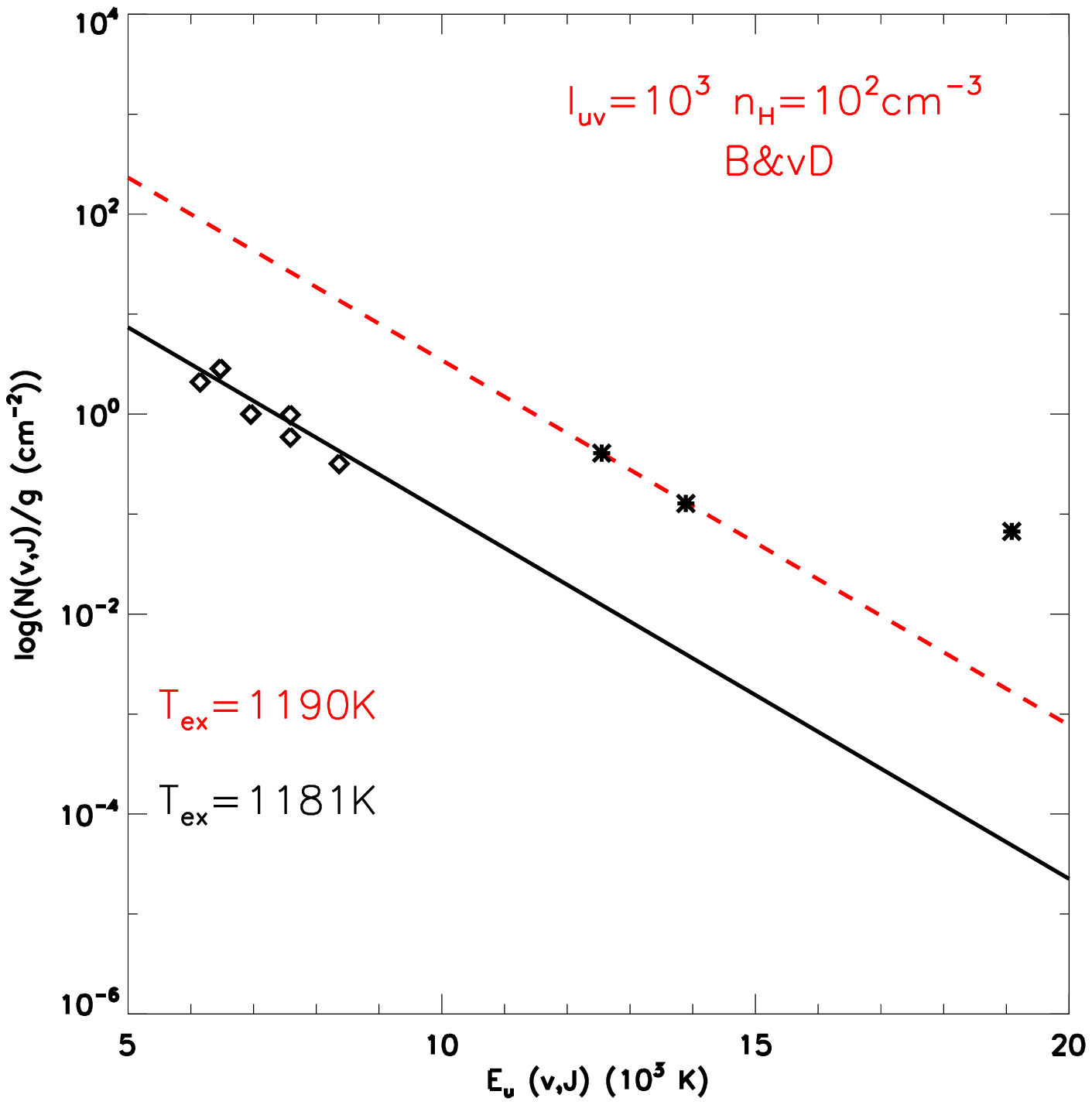}
\includegraphics[width=5.5cm]{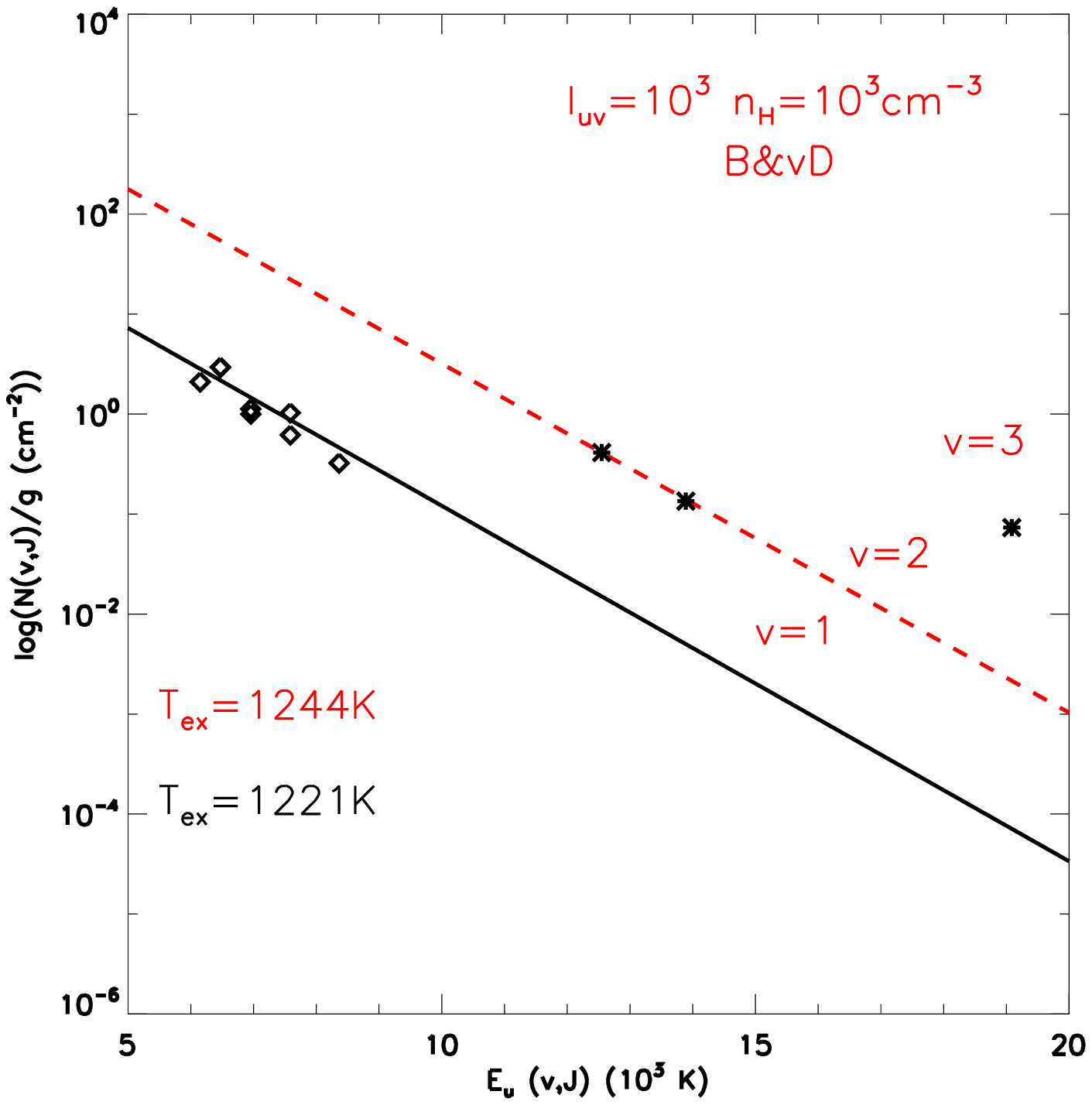}
\includegraphics[width=5.5cm]{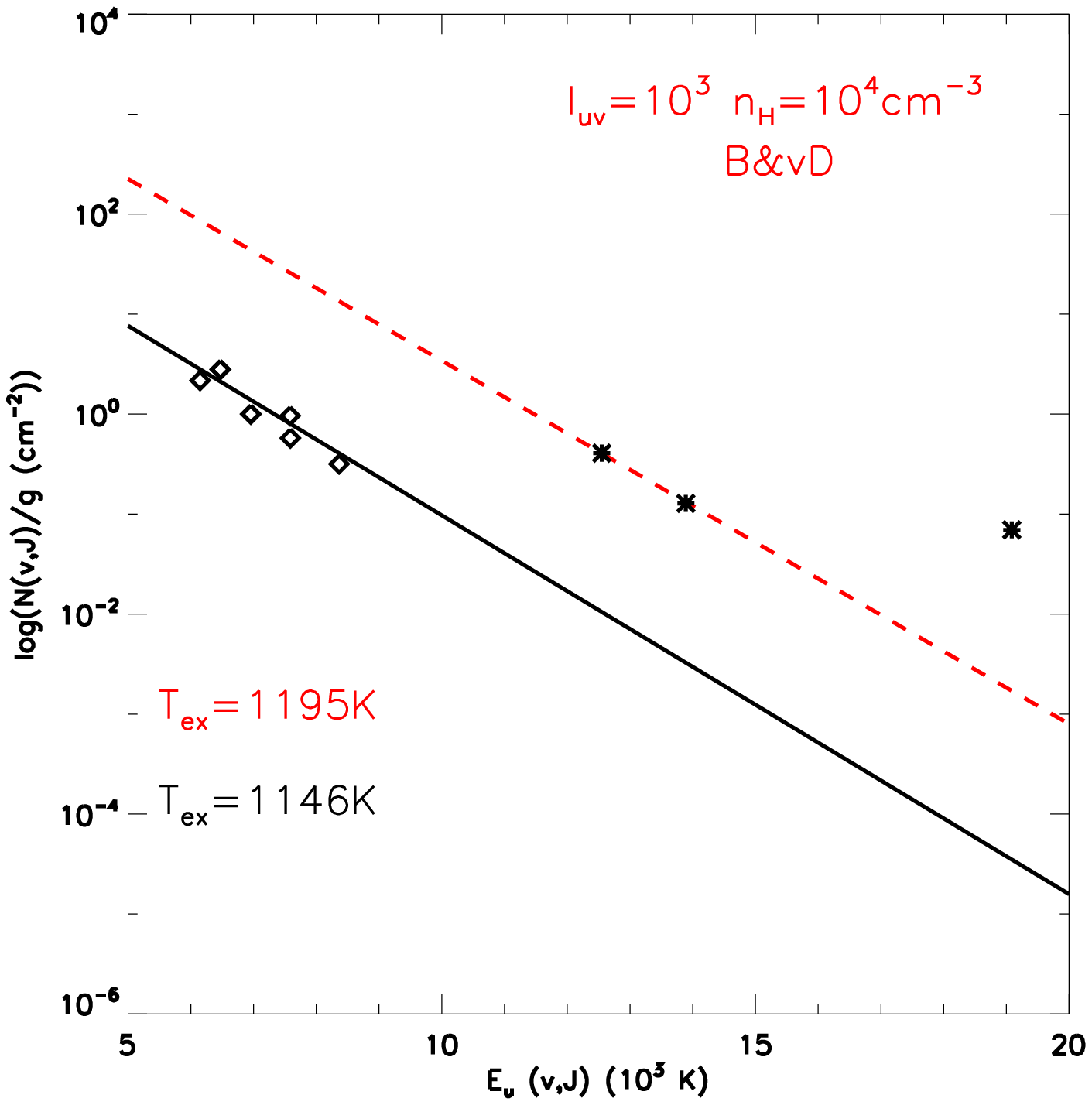}
\includegraphics[width=5.5cm]{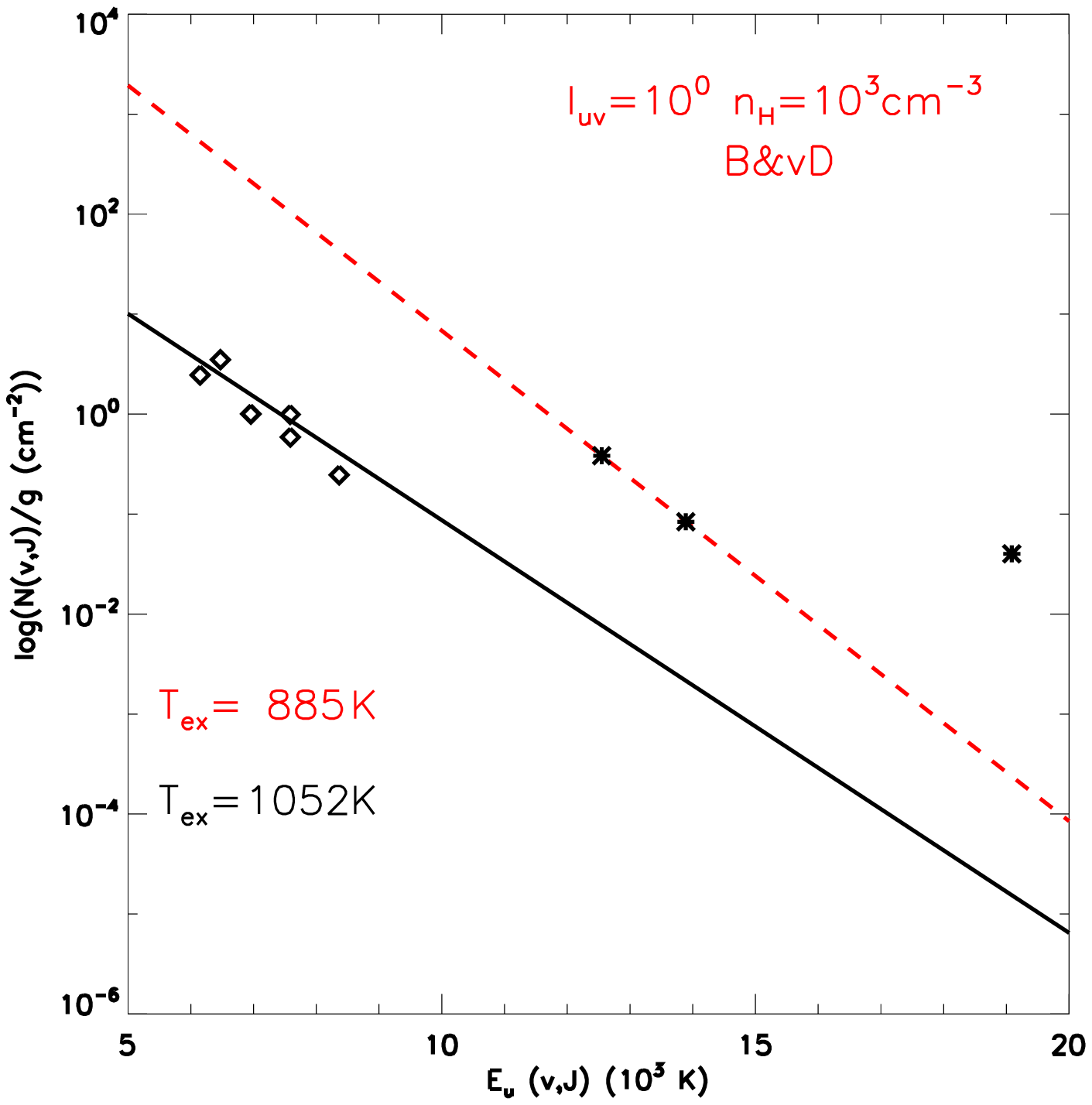}
\includegraphics[width=5.5cm]{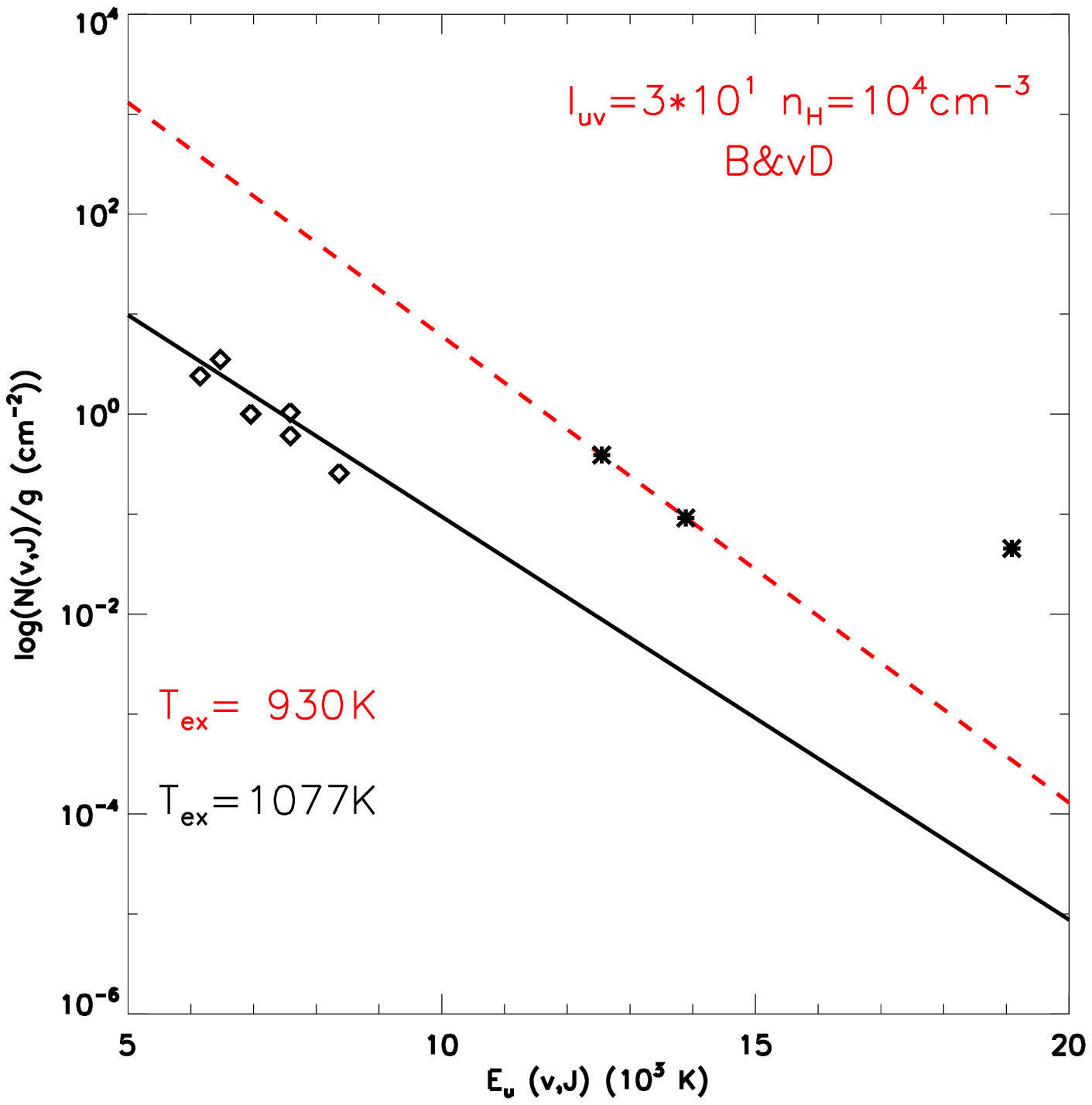}
\includegraphics[width=5.5cm]{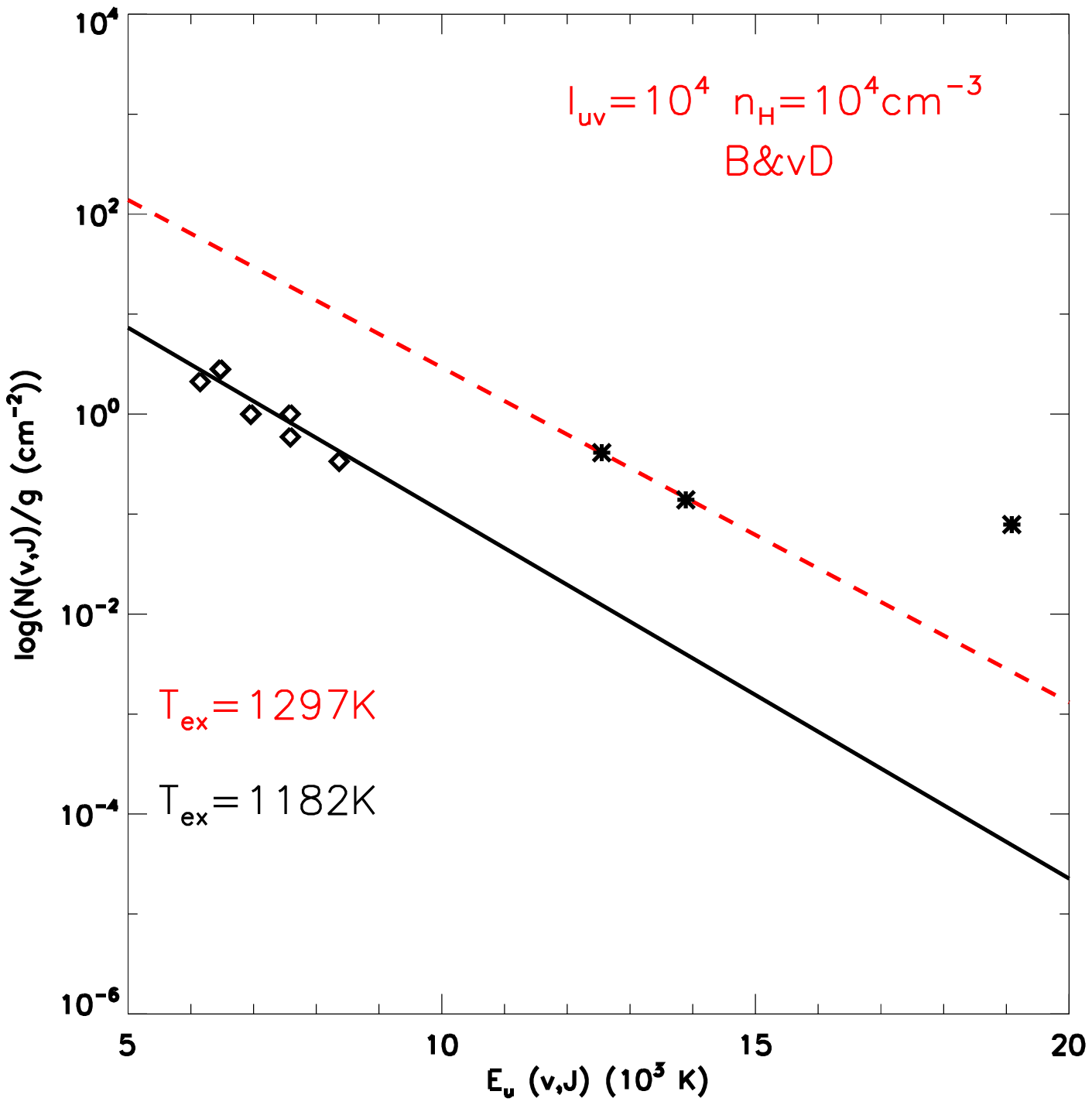}
\caption{Excitation diagrams of shock and PDR model-predicted H$_2$ line column density distributions as a function of upper energy level.  The column density is normalized to the 1-0 S(1) H$_2$ transition.  Diamonds represent lines from $v=1$ while the asterisks represent points from $v>1$.  The solid line is a best fit of the Boltzmann distribution to the $v=1$ points, the slope represents the excitation temperature, T$_{ex}$.  Predictions of shock models from \citet{1978ApJ...220..525S} with shock speeds 6 km/s (top row) and 14 km/s (second row) and densities n$_H$=$10^3$ cm$^{-3}$ (left), $10^4$ cm$^{-3}$(middle), and $10^5$ cm$^{-3}$ (right).  Predictions of PDR models from \citet{1987ApJ...322..412B} with I$_{uv}$=10$^0$-10$^4$ and densities n$_H$=$10^3$ cm$^{-3}$, $10^4$ cm$^{-3}$, and $10^5$ cm$^{-3}$, and are labeled with B$\&$vD.}
\label{fig:shockmod}
\end{figure*}

\begin{figure*}
\centering
\includegraphics[width=5.5cm]{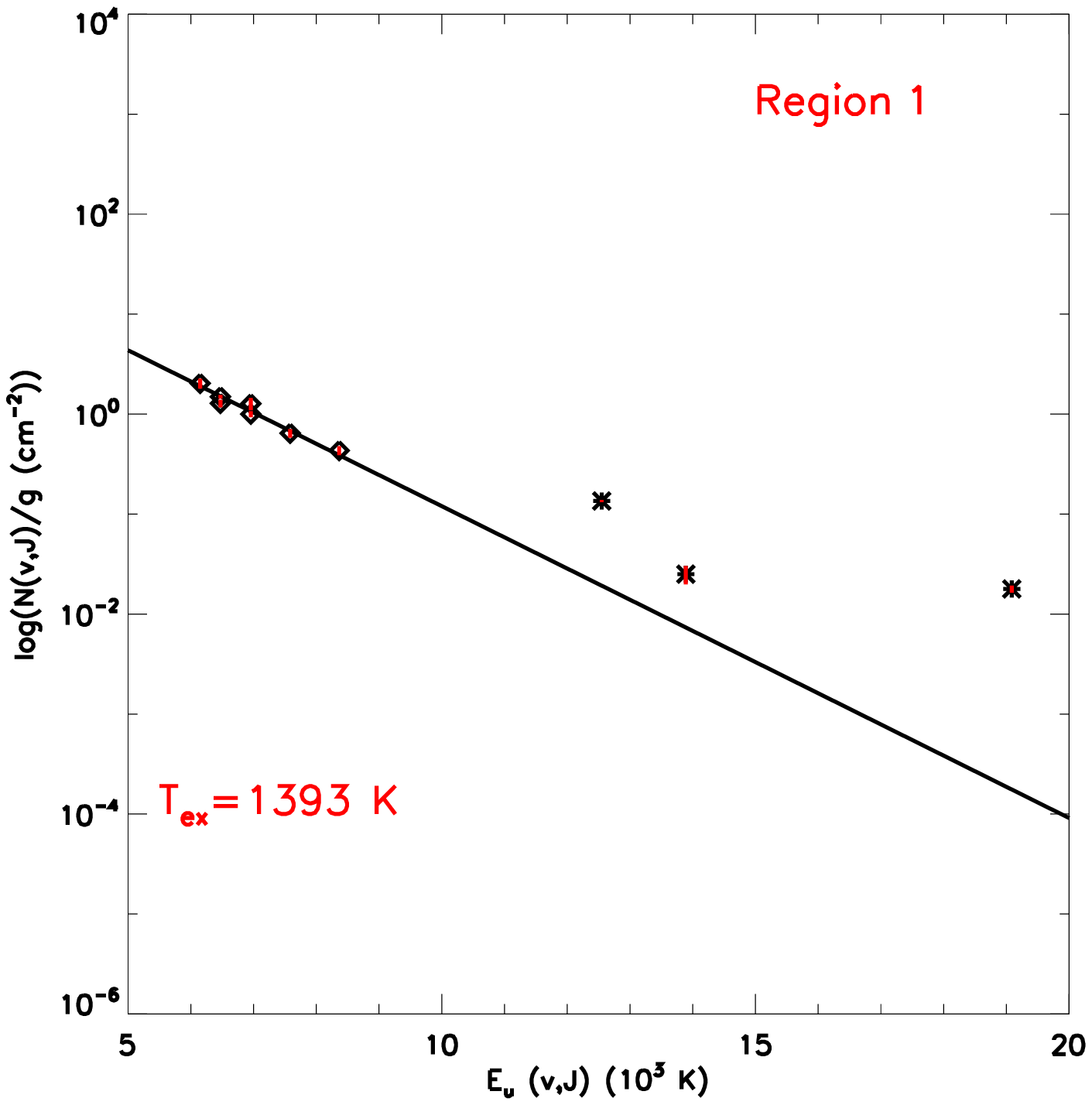}
\includegraphics[width=5.5cm]{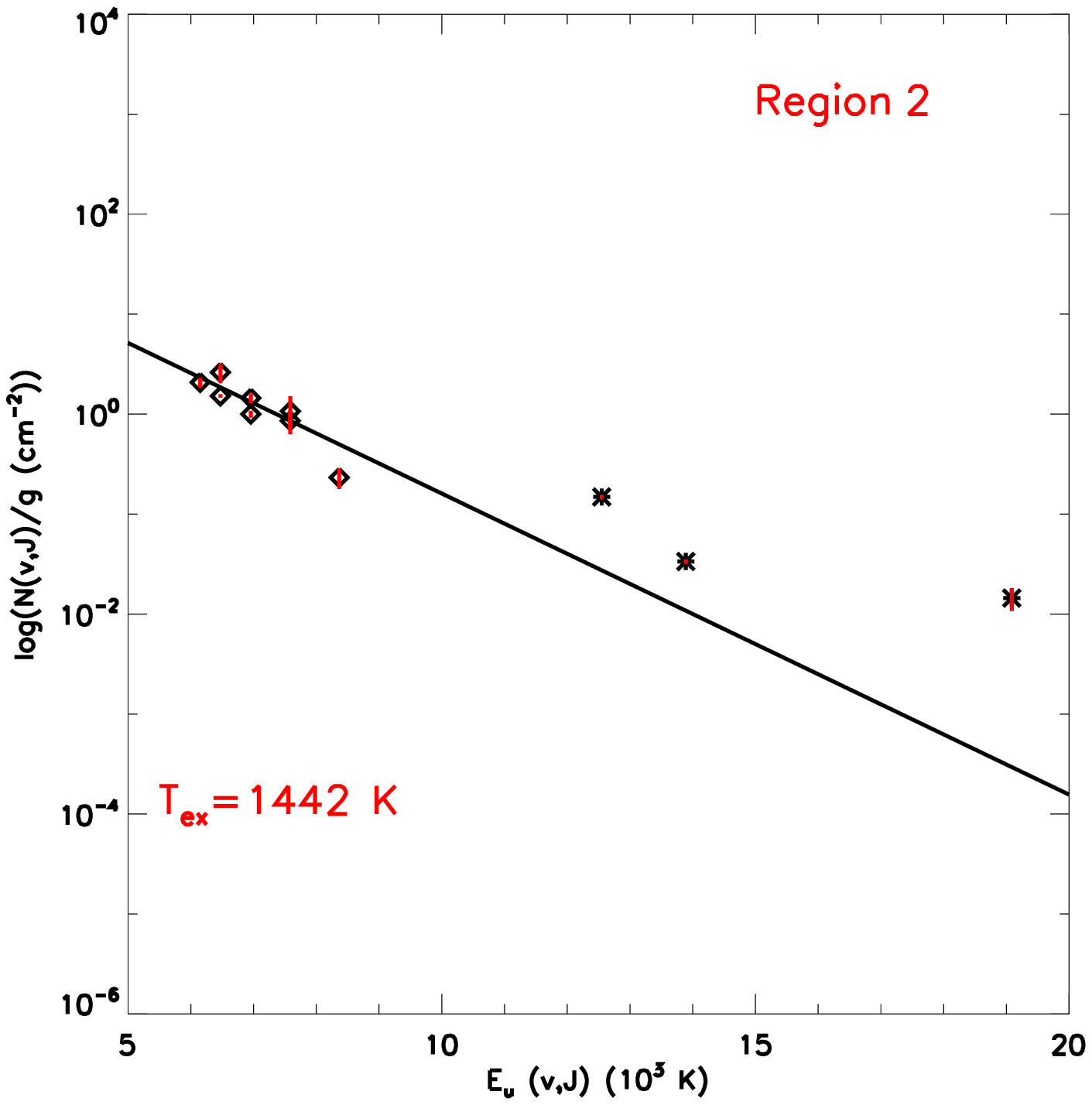}
\includegraphics[width=5.5cm]{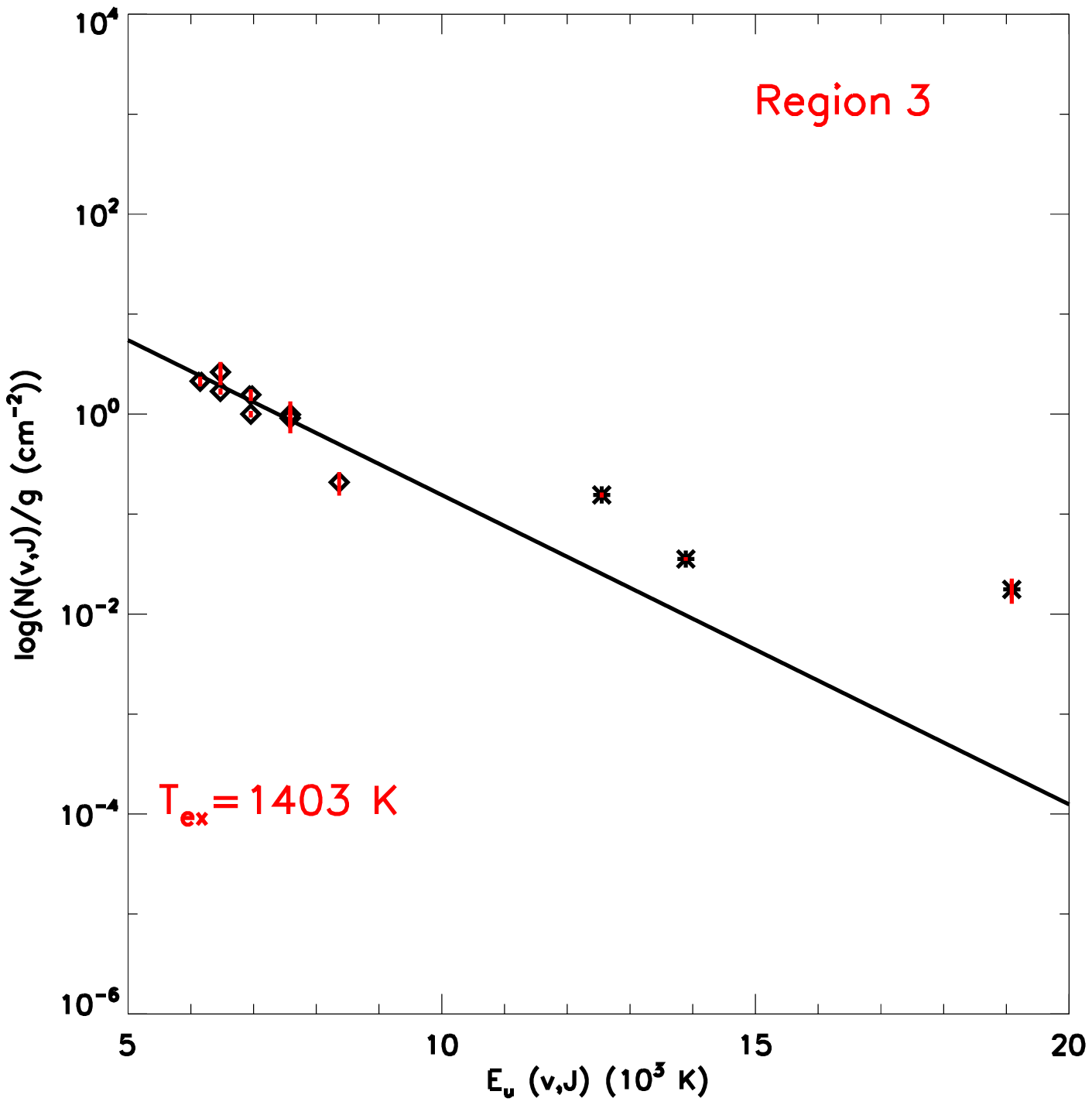}
\includegraphics[width=5.5cm]{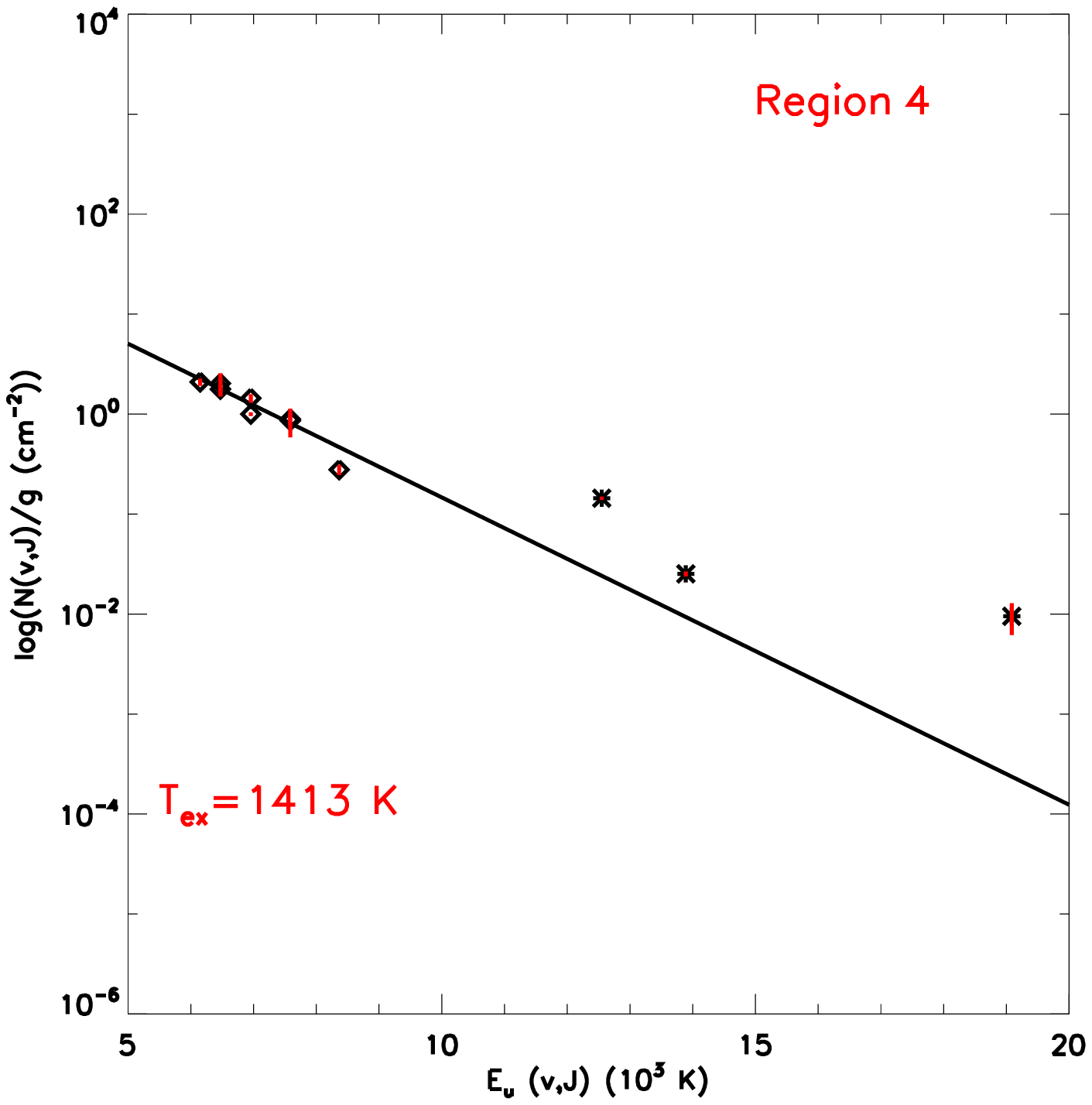}
\includegraphics[width=5.5cm]{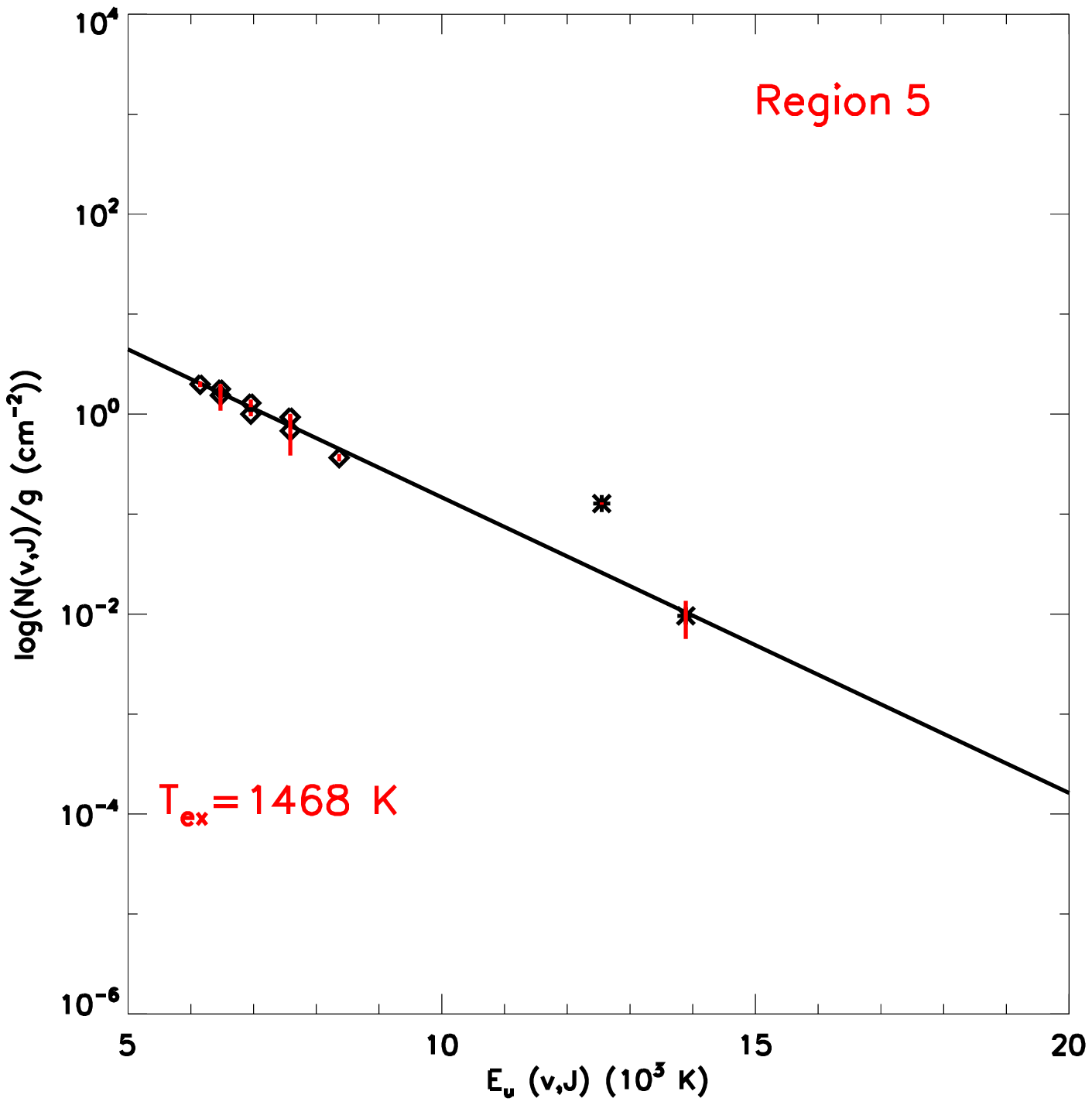}
\includegraphics[width=5.5cm]{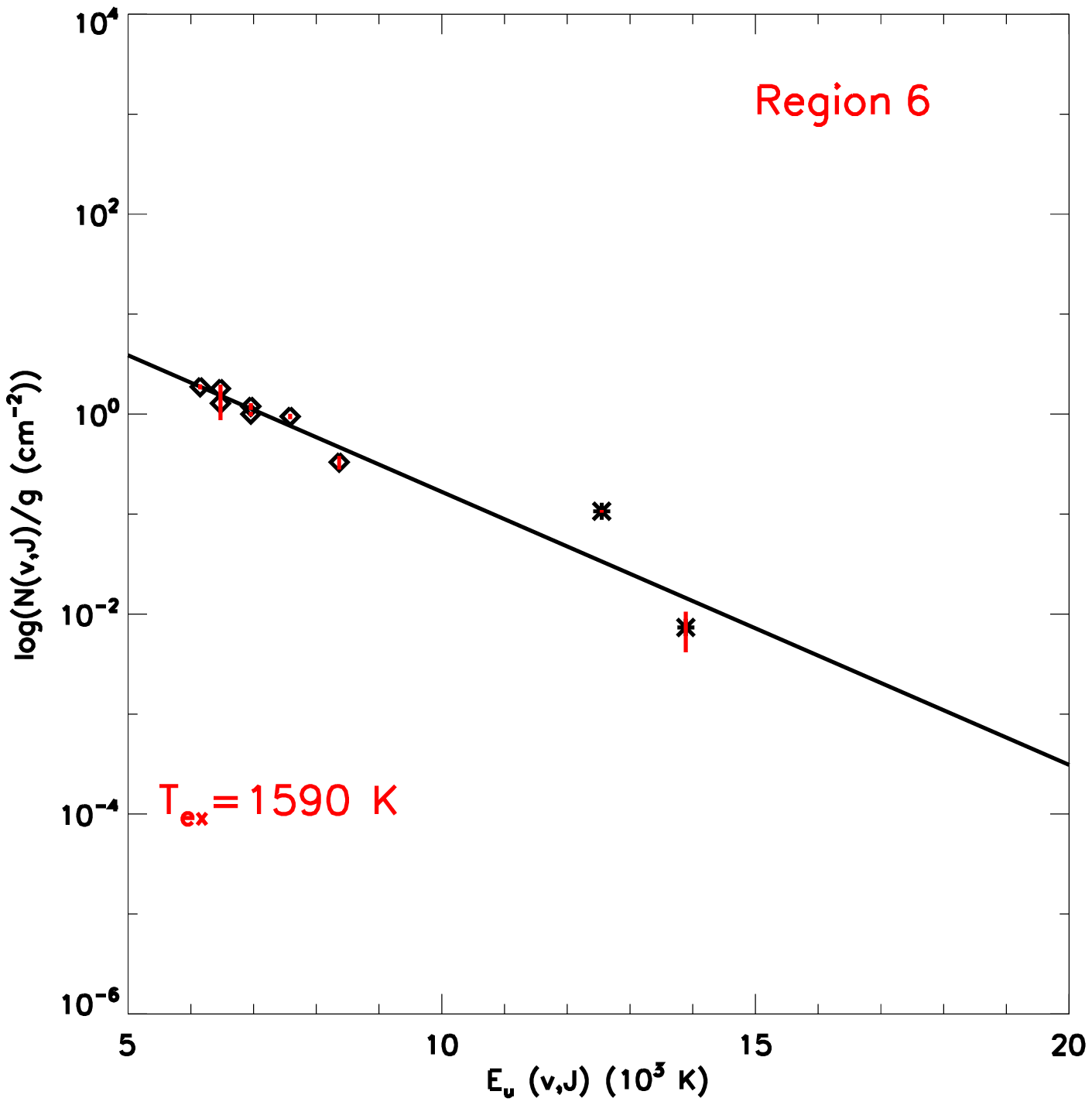}
\includegraphics[width=5.5cm]{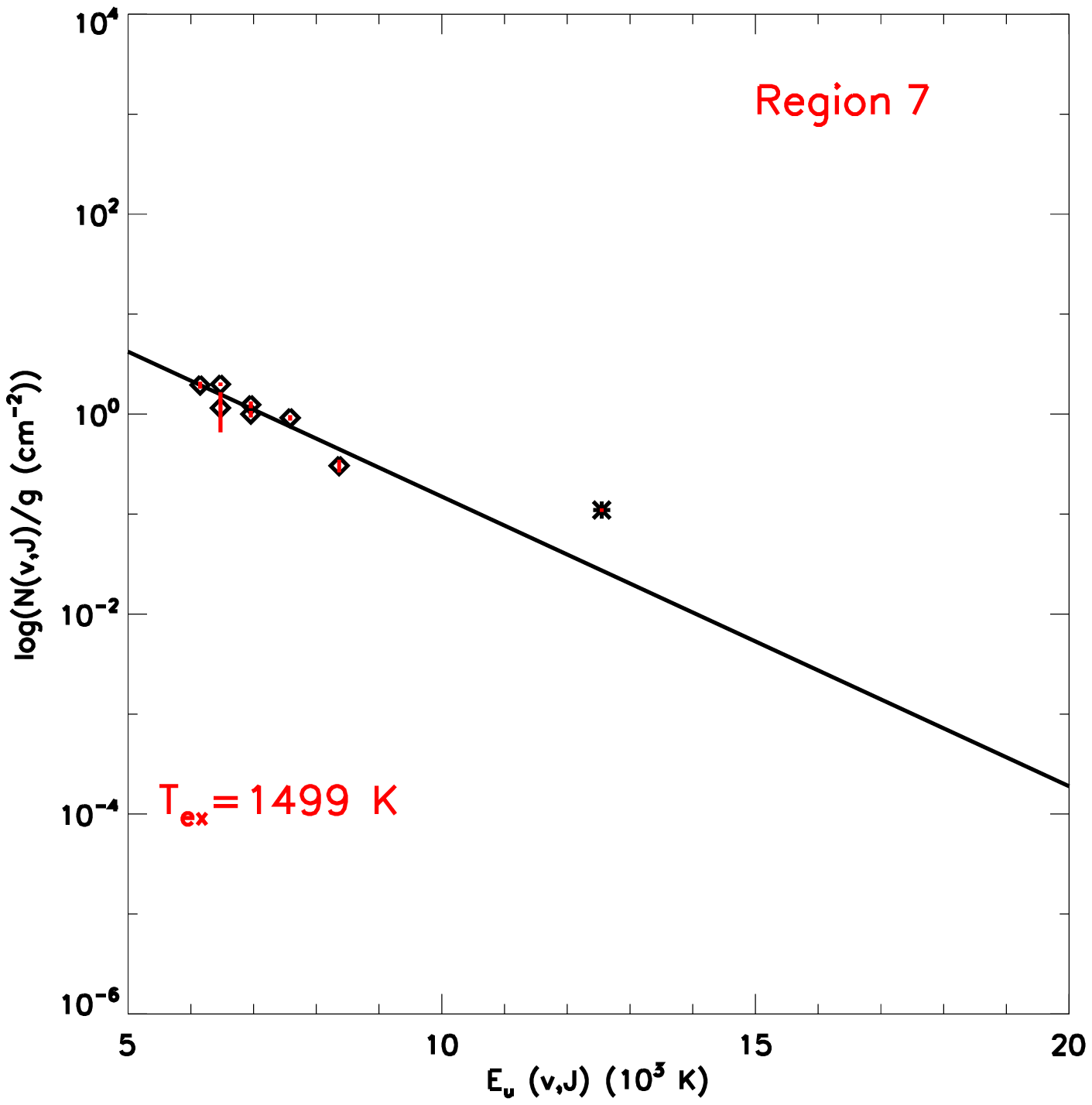}
\includegraphics[width=5.5cm]{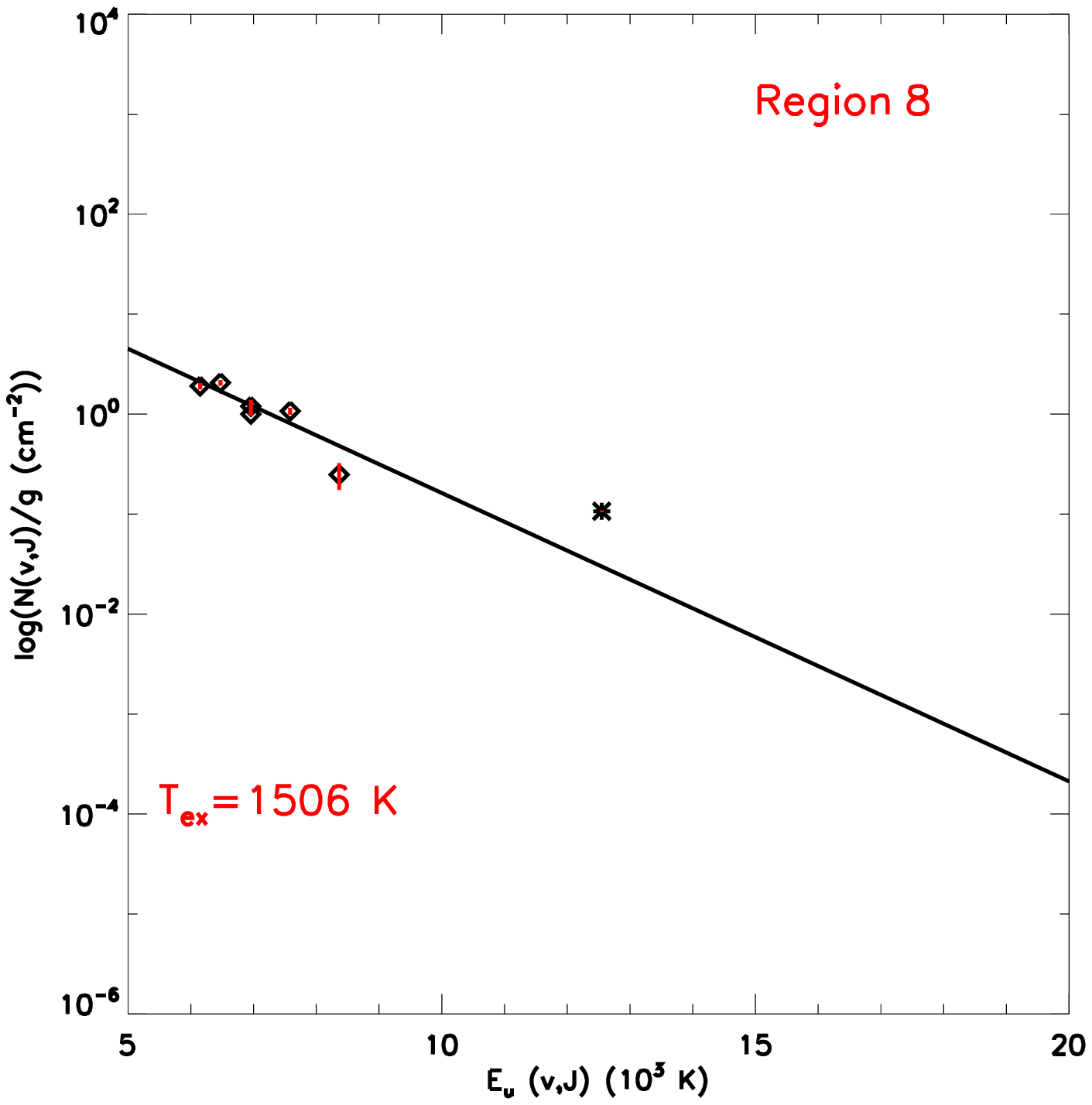}
\includegraphics[width=5.5cm]{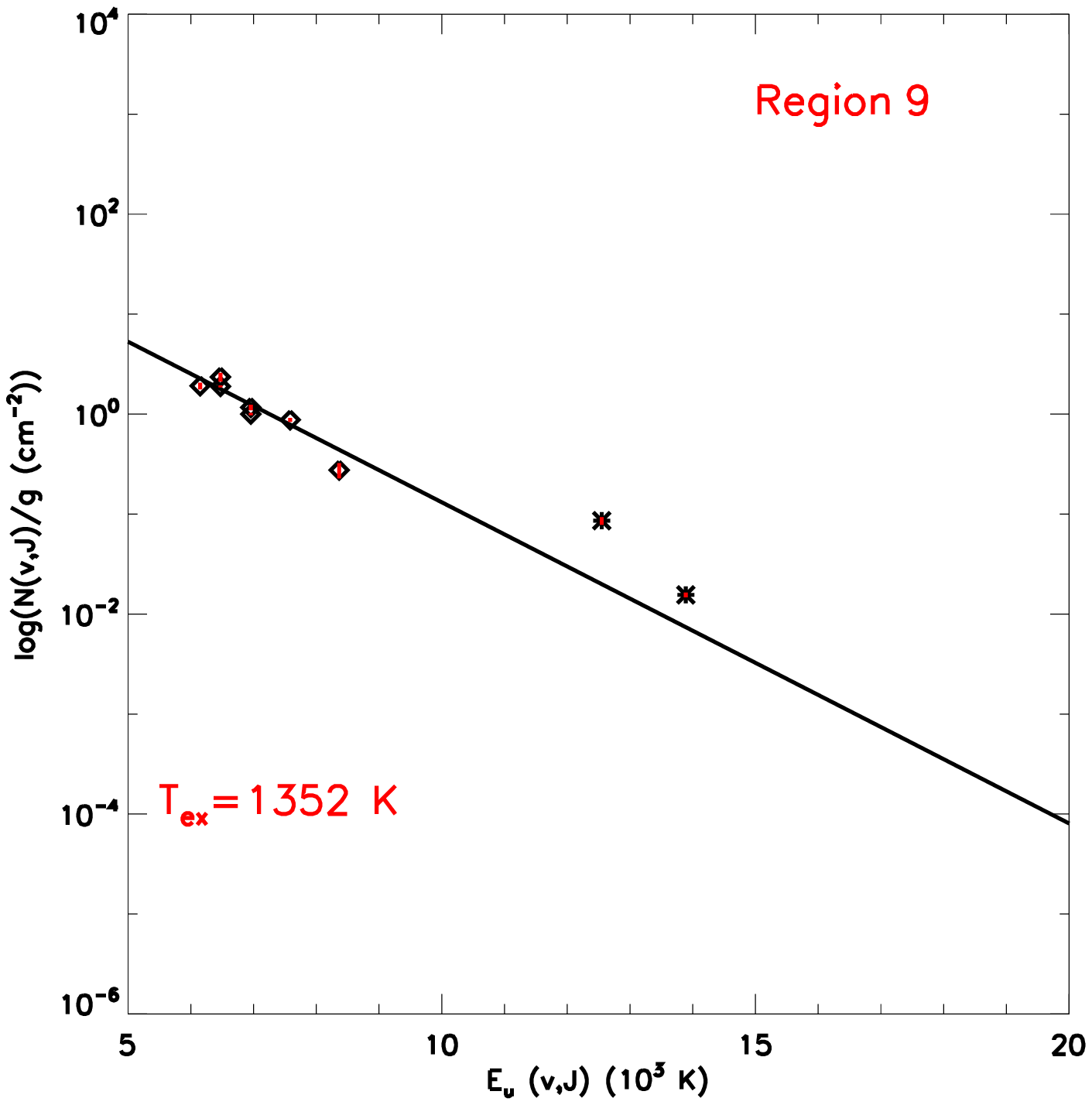}
\caption{Excitation diagrams of observed H$_2$ line column density distribution as a function of upper energy level.  The column density is normalized to the 1-0 S(1) H$_2$ transition.  Diamonds represent lines from $v=1$ while the asterisks represent points from $v>1$.  The solid line is a best fit of the Boltzmann distribution to the $v=1$ points, the slope represents the excitation temperature, T$_{ex}$.  Each panel represents a different region, corresponding to the regions in Figure~\ref{fig:h2ratio}.  The error bars represent a standard deviation from the column density.}
\label{fig:h2lines}
\end{figure*}

In order to directly compare the observed line ratios in NGC 253 to model predictions, we can use predicted line intensities for shock and PDR models and plot them on this diagnostic energy level diagram. Figure~\ref{fig:shockmod} presents the results from the \citet{1978ApJ...220..525S} shock models.  These models calculate line intensities for shocks moving at 6, 10, and 14 km/s for densities of 10$^3$, 10$^4$, $3\cdot10^5$ cm$^{-3}$.  In most of the shock models of Figure~\ref{fig:shockmod}, the $v>1$ lines (depicted as triangles) lie on or below the thermal distribution indicating subthermal excitation of the higher levels.  The exceptions to this case are the higher velocity (10-14 km/s) shocks in very dense environments (n$_H$=10$^5$).  However, the $v>1$ lines have a shallower slope than the thermal distribution as shown by the dotted red line, indicating a higher excitation temperature for these transitions.  In these high velocity, high density shocks, there are two different gas components (
temperature slopes).    

The bottom 6 panels of Figure~\ref{fig:shockmod} displays the predicted line intensities from \citet{1989ApJ...338..197S}, who model fluorescent excitation of H$_2$ in environments with n$_H=10^3-10^6$ and $\chi=10^2$ and \citet{1987ApJ...322..412B} who model pure fluorescent excitation for a large range of the parameter space.  We have chosen models 4, 6, 9, 17,19, and 20 spanning densities of n$_H=10^2-10^4$ cm$^{-3}$ and radiation intensities of $\chi=10^0-10^4$. In contrast to the shock models, the PDR models show the $v>1$ lines lying high above the thermal distribution in almost all cases, since UV fluorescence populates levels from the top down, the highest energy levels get populated with higher probability compared to the thermalized case.  In addition, the lines belonging to the $v=2$ level share the same excitation temperature (same slope) as the $v=1$ lines, only offset due to their higher upper energy level, unlike the shock models where there are two distinct excitation temperatures.  This is 
the same for each $v$ 
level, as 
shown in \citet{1987ApJ...322..412B}. A general feature of the UV-excited models is that the rotational temperatures (i.e., the excitation temperature determined between different rotational levels within the same vibrational level) are the same in the various vibrational levels. In contrast, the vibrational temperature (i.e., the excitation temperature determined between different vibrational levels) is higher than the rotational temperature. The only exception to this pattern is the top-right panel of Fig.~9, showing a model with low radiation field but very high density. However, this density is only relevant for very dense clumps and the the bulk of the molecular gas in NGC253 is at much lower densities.


Now that we have presented the shock and PDR models, we can compare them to the observational results seen in Figure~\ref{fig:h2lines}.  In these plots, only the ($\geq 2\sigma$) H$_2$ line detections are included.  Each panel of the figure represents the different regions shown in Figure~\ref{fig:kband}.  In each region, Figure~\ref{fig:h2lines} reveals the thermalized $v=1$ lines of varying excitation temperatures.  In addition, in all regions where $v>1$ lines are detected, they lie above the thermal distribution.  However, the case becomes ambiguous in regions where we cannot detect $v>1$ lines, such as Regions 8 and 9.  It is possible that in these regions shocks dominate and therefore the higher transitions are not being populated, or that in these regions, we lack the signal to noise to detect the lines above $2\sigma$. 

In both Region 1 and Region 3, the kinematic center, we detect two $v=2$ lines and one $v=3$ line.  Although we only have two $v=2$ lines, it is clear that they have a similar excitation temperature (slope) to the thermal distribution and do not represent a second hotter gas component.  Thus, these lines clearly follow a PDR distribution.  In addition, the $v>2$ lines cannot be excited through shocks, so the presence of the 3-2 S(3) line also suggests fluorescent excitation.  Using these criteria and the shock/PDR models as a guide, we can conclude that there is fluorescent excitation in Regions 1, 3, and 4 and most likely in Regions 2, 5 and 6.  Since the signal-to-noise is too low in Regions 7 and 8, we cannot determine whether shock excitation or fluorescent excitation is the dominant heating mechanism in these regions.  However, based on the integrated spectrum and Regions 1, 2, 3, and 4, it is clear that UV excitation in PDRs is the dominant mechanism 
for this galaxy, although shocks may dominate in some isolated areas.

\section{Conclusion }
\label{sec:conc}
From SINFONI observations, we determine the nature of the excitation of hot H$_2$ gas in NGC 253 using diagnostic emission lines.  Specifically, we use the K band continuum as a tracer of the older stellar population, Br$\gamma$ as a tracer of high-mass, young (O-type) star formation, [FeII] as a tracer of shocks, and PAHs as a tracer of slightly lower mass (B-type) star formation.  Based on the Br$\gamma$ and [FeII] emission line maps, we can interpret the H$_2$ excitation.  We find that in most regions of NGC 253, excitation by UV photons in PDRs is the dominant mechanism.  There are 3 regions where shock excitation may dominate, but these are small relatively isolated regions.  Throughout the entire nuclear region of the galaxy, H$_2$ is being fluorescently excited.  

We also present a diagnostic energy level diagram which robustly differentiates between shock and fluorescent excitation.  By comparing the observed column densities of each line to the proposed PDR and shock models, it is clear that the NGC 253 H$_2$ is dominated by fluorescent excitation. A few of the bright H$_2$ regions may lack the signal-to-noise to resolve the higher energy transitions, or it is possible that in these isolated regions, shocks are the dominant mechanism. In addition, the PAH emission follows the same morphology as the H$_2$, adding further confidence that through NGC 253, the gas is undergoing fluoresce.     

We determine that a maximum of 29\% of the hot molecular gas is excited by shocks.  Since we are only sensitive to the hottest H$_2$ gas, we are probing the surfaces of molecular clouds.  Within the molecular clouds, where the bulk of the molecular gas is, the gas can have a different excitation mechanism.

%
%
\begin{acknowledgements}
      We thank Lowell Tacconi-Garman for making his ISAAC 3.21 and 3.28 mu images available to us, and Susanne Brown for determining the calibration factors of these images.
\end{acknowledgements}

\bibliographystyle{aa}
\bibliography{bib_file.bib}
\listofobjects

\end{document}